\documentclass[article,nojss,shortnames]{jss}
\usepackage{amsmath}
\usepackage{amsfonts}
\usepackage{amsthm}
\usepackage{bm}
\usepackage{mleftright}
\usepackage{graphicx}
\usepackage{caption}
\usepackage{subcaption}
\usepackage{arydshln}
\usepackage{hyphenat}
\usepackage{colonequals}
\usepackage{booktabs}
\usepackage{lipsum}

\graphicspath{ {./figures/} }

\DeclareMathOperator*{\argmax}{argmax}

\definecolor{frenchblue}{rgb}{0.01,0.15,0.9}
\newcommand{\changed}[1]{{#1}}

\author{Jason Willwerscheid\\Providence College \And 
        Peter Carbonetto\\University of Chicago \And
        Matthew Stephens\\University of Chicago}
\title{\pkg{ebnm}: An \proglang{R} Package for Solving the Empirical Bayes Normal Means Problem Using a Variety of Prior Families}

\Plainauthor{Jason Willwerscheid, Peter Carbonetto, and Matthew Stephens} %
\Plaintitle{ebnm: An R Package for Solving the Empirical Bayes Normal Means Problem} %
\Shorttitle{\pkg{ebnm}: Empirical Bayes Normal Means in R} %

\Abstract{
The empirical Bayes normal means (EBNM) model
is important to many areas of statistics, 
including (but not limited to)
multiple testing, 
wavelet denoising, and
gene expression analysis.
There are several existing software 
packages that can fit EBNM models under different prior assumptions and using
different algorithms; however, the differences across interfaces complicate direct comparisons. Further, a number of important prior assumptions do not yet have implementations.
Motivated by these issues, we developed the \proglang{R} package \pkg{ebnm}, which provides a unified interface for efficiently fitting EBNM models 
using a variety of prior assumptions, including nonparametric approaches. 
In some cases, we incorporated existing implementations into \pkg{ebnm}; in others, we implemented new fitting procedures 
with a focus on speed and numerical stability.
\changed{
We illustrate the use of \pkg{ebnm} in a detailed analysis of baseball statistics. By providing a unified and easily extensible interface,
the \pkg{ebnm} package can facilitate 
development of new methods in statistics, genetics, and 
other areas; as an example, we briefly discuss the \proglang{R} package 
\pkg{flashier}, which harnesses methods in \pkg{ebnm} to provide a flexible and robust approach to matrix factorization. 
}
}


\Plainkeywords{empirical Bayes, normal means, shrinkage estimation, 
  mixture models, NPMLE, maximum likelihood} %

\Address{
  Jason Willwerscheid\\
  Department of Mathematics and Computer Science\\
  Providence College\\
  Providence, Rhode Island, United States of America\\
  E-mail: \email{jwillwer@providence.edu}\\

\medskip
  Peter Carbonetto\\
  Department of Human Genetics and the
  Research Computing Center\\
  University of Chicago\\
  Chicago, Illinois, United States of America\\
  E-mail: \email{pcarbo@uchicago.edu}\\

\medskip

  Matthew Stephens\\
  Departments of Statistics and Human Genetics\\
  University of Chicago\\
  Chicago, Illinois, United States of America\\
  E-mail: \email{mstephens@uchicago.edu}
}

\def\G{\mathcal{G}}

\begin{document}

\section{Introduction}
\label{section:intro}

Given $n$ observations $x_i \in \mathbb{R}$ with known standard
deviations $s_i > 0$, $i = 1, \dots, n$, the normal means model
\citep{Robbins51, efron1972limiting, Stephens_NewDeal,
  bhadra2019lasso, Johnstone, lei-thesis} has
\begin{equation}
x_i \overset{\text{ind.}}{\sim} \mathcal{N}(\theta_i, s_i^2),
\label{eqn:nm_problem}
\end{equation}
where the unknown (``true'') means $\theta_i \in \mathbb{R}$ are the 
quantities to be
estimated. Here and throughout, we use $\mathcal{N}(\mu, \sigma^2)$ to
denote the normal distribution with mean $\mu$ and variance
$\sigma^2$. %

\begin{figure}[!t]
\centering
\includegraphics[width=\textwidth]{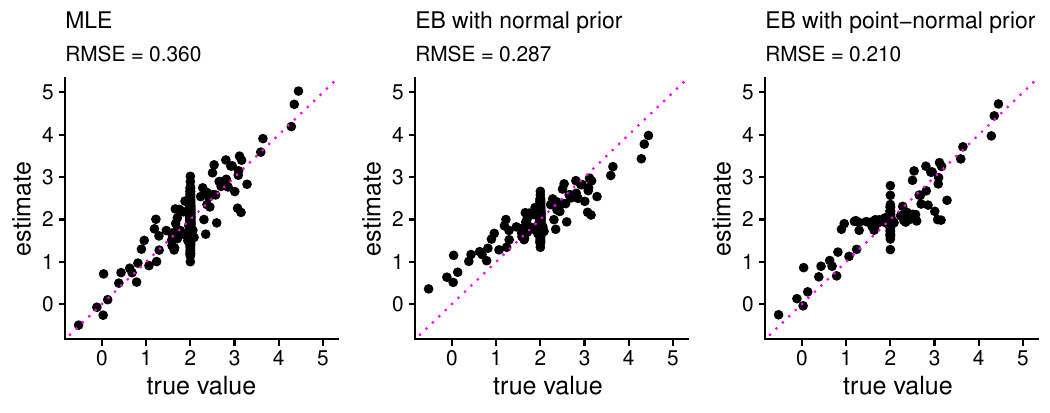}
\caption{\changed{Illustration of shrinkage estimation using empirical Bayes normal
means. The example data set consists of 400 noisy observations
$x_i$, with true means
$\theta_i$ simulated from a ``point-normal'' prior (see
Table~\ref{table:ebnm_prior_families}). The left-hand plot shows the observations $x_i$ (which are also
the MLE estimates of the true means) as a function of the
true means $\theta_i$. The middle plot shows EB
estimates of the true means obtained by learning a normal prior $g$
from the data. These EB estimates ``shrink'' the observations toward the mode of the common prior, improving the overall
root mean-squared error (RMSE) from 0.360 to 0.287. But the normal
prior also appears to ``overshrink'' observations distant from the center
(near 2). Fitting an EBNM model with a family of priors that
better suits the data --- that is, point-normal priors --- results in more
accurate EB estimates, improving the overall RMSE
to 0.210. In particular, it avoids ``overshrinking'' more extreme observations; see the right-hand plot. For an expanded
illustration, including the \proglang{R} code that reproduces the results shown
here, see the \pkg{ebnm} package vignette, ``Introduction to the empirical Bayes normal means model via shrinkage estimation.''}}
\label{fig:shrink_intro}
\end{figure}

The empirical Bayes (EB) approach to inferring
$\theta_i$ --- which we refer to as ``solving the empirical Bayes normal means problem'' --- attempts to improve
upon the maximum-likelihood estimate $\hat{\theta}_i = x_i$ by
``borrowing information'' across observations, exploiting the fact
that each observation contains information not only about its
respective mean, but also about how the means are collectively
distributed \citep{Robbins56, Morris, Efron_Book, Stephens_NewDeal}.
Specifically, the empirical Bayes normal means (EBNM) approach assumes
that
\begin{equation} \label{eqn:nm_prior}
\theta_i \overset{\text{ind.}}{\sim} g \in \mathcal{G},
\end{equation}
where $\mathcal{G}$ is some family of probability distributions that is specified
in advance and $g \in \mathcal{G}$ is estimated using the data
(typically, via maximum likelihood). Given $\hat{g} \in \mathcal{G}$, estimates of $\theta_i$ can be obtained using, for example,  posterior means:
\begin{equation}
    \hat{\theta}_i := \mathbb{E} \left( \theta_i \mid x_i, \hat{g} \right).
\end{equation}
\changed{See Figure~\ref{fig:shrink_intro} for an illustration.}

Applications in which the EBNM model plays an important role include wavelet
denoising \citep{ClydeGeorge,JohnstoneSilverman_Needles,
  JohnstoneSilverman_Wavelet}; multiple testing \citep{Efron_Book,
  Stephens_NewDeal}; gene expression analysis \citep{deseq2, zhu-2019,
  smyth2004linear}; multiple linear regression \citep{mr_ash,
  mukherjee-2023}; and matrix factorization \citep{WangStephens}.

This versatility has motivated the development of
a number of software packages using different choices of prior family $\mathcal{G}$; for a review, see Section \ref{section:ebnm_background}. Still, important gaps in the software
remain. For example, we are not aware of any package that fits the
EBNM model in the simple case where $\mathcal{G}$ is taken to be the
family of all univariate normal distributions.
Further, each existing package has a different interface and
outputs, which hinders comparisons as well as making it
difficult to develop software packages that flexibly build upon EBNM methods.
Motivated by these issues, we developed the 
\pkg{ebnm} software package, which provides a unified interface for efficiently solving
the EBNM problem using a wide variety of prior families. 

\changed{
We wrote the \pkg{ebnm} package in \proglang{R} \citep{r}, which is free, open source, and highly interoperable --- for example, with \proglang{Python} via package \pkg{rpy2} \citep{rpy2}; with \proglang{MATLAB} via \pkg{R-link} \citep{rlink}; and with \proglang{Julia} via \pkg{RCall} \citep{rcall}. The \pkg{ebnm} package can be downloaded from the Comprehensive R Archive Network (CRAN), and the latest development branch is available on GitHub (\url{https://github.com/stephenslab/ebnm}). The package's website, which includes detailed documentation and package vignettes, is available at \url{https://stephenslab.github.io/ebnm/}. Code for reproducing results and figures in the text is available on GitHub at 
\url{https://github.com/willwerscheid/ebnm-paper}.}

The paper is organized as follows. In Section~\ref{section:ebnm_background}, we give a brief history of
the EBNM problem and review existing approaches. Section~\ref{section:ebnm_methods} gives an overview of the \pkg{ebnm}
package, including the unified interface and the newly implemented
prior families. In Section~\ref{section:ebnm_simdata}, we compare
different choices of prior family and illustrate how this choice of prior family can impact
performance. 
This section also includes a runtime benchmark for implemented prior families. 
\changed{
Section~\ref{section:woba} illustrates usage of the \pkg{ebnm} package in
an
analysis of baseball statistics. In Section~\ref{section:ebnm_snmf}, we outline a matrix factorization approach, implemented in the \proglang{R} package \pkg{flashier}, which builds on the methods in \pkg{ebnm}. Here, the EBNM problem arises as a subproblem that must be solved many times, so that the speed and flexibility of the \pkg{ebnm} methods prove
critical. Finally, Section \ref{sec:summary} summarizes the key contributions 
of this work.}

\section{Background and existing software}
\label{section:ebnm_background}

In this section we review existing approaches to the EBNM problem within a common
modeling framework.

\subsection{Normal priors}

\cite{Stein} famously discovered that under quadratic loss, the
maximum-likelihood estimate (MLE) $\hat{\theta}_i = x_i$, $i = 1, \ldots,
n$, is an inadmissible solution to the homoskedastic normal means
problem
\begin{equation} \label{eqn:homosked}
x_i \overset{\text{ind.}}{\sim} \mathcal{N}(\theta_i, s^2),
\qquad i=1,\ldots,n,
\end{equation}
when $n \geq 3$. \cite{JamesStein} subsequently gave an explicit
formula for a shrinkage estimator that dominates the
MLE. As \cite{EfronMorris} showed, a
lightly modified version of the James-Stein estimator can be derived
via an EB approach that assumes
\begin{equation} \label{eqn:nm_prior_homoskd}
\theta_i \overset{\text{ind.}}{\sim} g \in \mathcal{G},
\end{equation}
where the prior family $\mathcal{G}$ is taken to be the family of
zero-mean normal distributions,
\begin{equation}
\mathcal{G}_{\text{norm0}} \colonequals 
\big\{g: g(x) = \mathcal{N} (x; 0, \sigma^2),\,
\sigma^2 \ge 0 \big\}.
\label{eqn:G_norm0}
\end{equation}
Here, we use $\mathcal{N}(x; \mu, \sigma^2)$ to denote the normal
probability density function at $x$ with mean $\mu$ and variance
$\sigma^2$.

In many applications, the mean of the $\theta_i$'s may be non-zero, and
so a natural generalization takes $\mathcal{G}$ to be the family of
normal distributions
\begin{equation}
\mathcal{G}_{\text{norm}} \colonequals \big\{g: g(x) = 
\mathcal{N}(x; \mu, \sigma^2),\, \sigma^2 \ge 0
\mbox{ and }
\mu \in \mathbb{R} \big\}.
\label{eqn:G_norm}
\end{equation}
Estimating $g \in \G_{\text{norm}}$ reduces to estimating $\sigma^2$
and $\mu$. For the homoskedastic case
\eqref{eqn:homosked},
the maximum-likelihood estimates have simple, closed-form solutions:
\begin{align} 
\hat{\mu} &= \frac{1}{n} \sum_{i=1}^n x_i, \\
\hat{\sigma}^2 &= 
\max \bigg\{ 0, \frac{1}{n} \sum_{i=1}^n (x_i-\hat{\mu})^2 - s^2 \bigg\}.
\end{align}
When $\mu$ is fixed at zero, this solution is similar to the one
implied by the positive-part James-Stein estimator, with the
difference that it divides $\sum_{i=1}^n x_i^2$ by $n$ rather than by
$n - 2$ \citep{EfronMorris}. For the heteroskedastic case
\eqref{eqn:nm_problem}, the likelihood $L(\mu,\sigma^2)$ has a closed
form but must be maximized numerically. 

In both the homoskedastic and
heteroskedastic cases, the posterior distributions
\begin{equation} 
\label{eqn:post}
p(\theta_i \mid x_i, s_i, \hat{g}) \propto 
\hat{g}(\theta_i) \, p(x_i \mid \theta_i, s_i)
\end{equation}
are normal
distributions which are available analytically. 

\subsection{Sparse priors}

Although the normal prior family has the advantage of simplicity, in practice more flexible priors are usually preferred. In particular, one would often like the
prior to be able to capture sparsity in $\boldsymbol{\theta} := \{\theta_1, \ldots, \theta_n\}$. One approach is to
use a ``spike-and-slab'' prior; that is, a mixture consisting of two
components, one a point-mass at zero (the ``spike'') and the other
(the ``slab'') belonging to some family of continuous
distributions, usually symmetric and centered at zero. A common choice
is the ``point-normal'' family,
\begin{equation}
\mathcal{G}_{\text{pn}} \colonequals
\big\{g: g(x) = \pi_0 \delta_0(x) + (1 - \pi_0) 
\mathcal{N}(x; 0, \sigma^2), \,
0 \le \pi_0 \le 1, \,
\sigma^2 > 0 \big\},
\label{eqn:G_pn}
\end{equation}
in which $\delta_y(x)$ denotes the density function at $x$ for the
delta-Dirac mass centered at $y$. With this choice, estimating $g$
reduces to estimating two parameters, $\pi_0$ and $\sigma$. Similar to
the family of normal distributions, the likelihood
for the point-normal prior family has a closed form, and standard numerical
optimization methods can be used to efficiently find the
maximum-likelihood solution. 
Given 
$\hat{g} \in \G_\text{pn}$, the posterior distributions \eqref{eqn:post} are
mixtures of a point-mass at zero and a normal distribution, and are
available analytically. 

As \cite{JohnstoneSilverman_Wavelet} showed,
replacing the normal slab with a ``heavy-tailed'' distribution
generally improves accuracy. Their \pkg{EbayesThresh} software,
available in \proglang{R} and \proglang{S-PLUS}
\citep{EbayesThresh_paper}, implements two such
priors: the point-Laplace prior,
\begin{equation} 
\mathcal{G}_{\text{pl}} \colonequals 
\big\{g: g(x) = \pi_0 \delta_0(x) + 
(1 - \pi_0) \mathrm{Laplace}(x; 0, a), \,
0 \le \pi_0 \le 1, \,
a \ge 0 \big\},
\label{eqn:G_pl}
\end{equation}
in which $\mathrm{Laplace}(x; \mu, a)$ denotes the probability density of
the Laplace distribution \citep{GelmanBDA} at $x$ with mean $\mu$ and
scale $a$, and a family of priors in which the slab has
``Cauchy-like'' tails. Again,
maximum-likelihood estimates $\hat{g} \in \mathcal{G}$ can be found using numerical methods.

Another parametric prior that is well-suited for capturing sparse signals is the horseshoe prior
\citep{Horseshoe},
which models sparsity by having appreciable mass near zero rather than
exactly at zero. The \proglang{R} package \pkg{horseshoe}
\citep{Horseshoe_Package} solves the
homoskedastic EBNM problem with $\G$ the family of horseshoe
distributions. See \cite{bhadra2019lasso} for a review of other implementations of the horseshoe prior.

\subsection{Nonparametric approaches}
\label{section:nonparametric}

The estimate of $g$ when $\mathcal{G}$ is the unconstrained family of {\em all}
distributions is called the nonparametric maximum-likelihood estimate
(NPMLE) \citep{KieferWolfowitz, Laird, Lindsay,
  JiangZhang, KoenkerMizera, DickerZhao}. In practice, most
nonparametric methods approximate this family, which we denote as $\mathcal{G}_{\text{npmle}}$, by
a dense but finite mixture of point masses,
\begin{equation}
\tilde{\mathcal{G}}_\text{npmle} 
\colonequals \bigg\{ g: g(x) = \sum_{k=1}^K \pi_k \delta_{\mu_k}(x) 
\,\bigg|\, \pi_1, \ldots, \pi_K \ge 0,\, 
\sum_{k = 1}^K \pi_k = 1 \bigg\},
\label{eq:g_npmle}
\end{equation}
where $\mu_{1}, \dots, \mu_K$ is a fixed, dense grid of values spanning the range of the observations. Estimating $g \in
\tilde{\mathcal{G}}_{\text{npmle}}$ amounts to solving the 
constrained optimization problem,
\begin{equation} 
\label{eqn:opt_problem}
\begin{array}{ll}
\mbox{maximize} &\mathbf{L} \boldsymbol{\pi} \\
\mbox{subject to} &\boldsymbol{0} \preceq 
\boldsymbol{\pi} \preceq \boldsymbol{1}_K \\
&\boldsymbol{\pi}^\top \boldsymbol{1}_K = 1,
\end{array}
\end{equation}
where ${\bm\pi} \colonequals (\pi_1, \ldots, \pi_K)^\top$, ${\bm 1}_K$ is
a column vector of ones of length $K$, and $\mathbf{L} \in
\mathbb{R}^{n \times K}$ is the matrix with entries 
$\ell_{ik} = \mathcal{N}(x_i; \mu_k, \sigma_i^2)$.
This is a convex optimization problem \citep{KoenkerMizera}. The
\proglang{R} package \pkg{REBayes} \citep{KoenkerGu} implements an
efficient solution based on interior point optimization methods
\citep{mosek}, but see \cite{MixSQP} and \cite{ZhangCui} for other approaches.

Although the fully nonparametric approach is the most flexible, the NPMLE is a discrete distribution \citep{Laird}, so posterior distributions \eqref{eqn:post} are discrete as well. While the posterior
mean from a discrete prior may be perfectly adequate for point
estimation, %
interval estimates can be unsatisfactory.  
The \proglang{R} package \pkg{deconvolveR} \citep{NarasimhanEfron}
uses a natural spline basis to obtain a smoothed nonparametric
estimate of $g$, which yields sensible interval estimates and can outperform the NPMLE in certain respects when the true prior is smooth 
\citep{KoenkerVinaigrette}.

\subsection{Constrained nonparametric approaches}
\label{section:constrained_nonpar}

Constrained nonparametric approaches can offer a middle ground, retaining much of the flexibility of fully nonparametric approaches while avoiding the potential perils of ``overfitting'' \citep{hastie2009elements}.
\cite{Stephens_NewDeal} argued that the set of
all distributions that are unimodal at zero can be a particularly good choice for
$\mathcal{G}$ in the context
of multiple testing. If it is reasonable to assume that the prior $g$ is symmetric, one can instead take $\mathcal{G}$ to be the 
family of scale
mixtures of normals,
\begin{equation} 
\mathcal{G}_{\text{smn}} \colonequals \textstyle 
\big\{g: g(x) = \int_0^{\infty} \mathcal{N}(x; 0,\sigma^2) \, dh 
(\sigma^2) \text{ for some } h \big\}, 
\label{eqn:G_smn}
\end{equation}
or, for slightly more flexibility, the family of all symmetric distributions that are unimodal at zero, which can be represented by 
scale mixtures of uniform distributions,
\begin{equation} 
\mathcal{G}_{\text{symm-u}} \colonequals \textstyle
\big\{g: g(x) = \int_0^{\infty} \mathrm{Unif}(x; -a, a) \, dh
\left( a \right) \text{ for some } 
h \big\},\label{eqn:G_symm}
\end{equation}
in which $\mathrm{Unif}(x; a, b)$ denotes the probability density
function at $x$ of the uniform distribution on the interval $[a, b]$.

When these families are approximated by finite mixtures, estimating
$g$ reduces to the same convex optimization problem that arises from the NPMLE, and can again be
solved using fast algorithms for convex optimization. This approach is implemented in the \proglang{R} package
\pkg{ashr} \citep{ashr}, which, by default, uses \pkg{mixsqp}
\citep{MixSQP} to solve the optimization problem \eqref{eqn:opt_problem}.

\section[The ebnm package: implementation and usage]{The 
\pkg{ebnm} package: implementation and usage}
\label{section:ebnm_methods}

The \pkg{ebnm} package provides a unified interface for solving the EBNM
problem using a wide variety of prior assumptions, including all of the choices of prior family discussed above in Section \ref{section:ebnm_background}. In addition to making available existing implementations via a shared
interface, the package provides new
implementations for several simple but
useful prior families that, to our knowledge, have not previously been
implemented, such as the normal and point-normal prior families $\mathcal{G}_{\text{norm}}$ and $\mathcal{G}_{\text{pn}}$. 
\changed{The underlying computations for all prior families have been carefully optimized 
(see Section \ref{section:ebnm_opt} below).
The implemented prior families are summarized in Table~\ref{table:ebnm_prior_families}. Note, also, that
\pkg{ebnm} was designed to be easily extensible 
to other prior families; to facilitate and encourage such extensions, we have written a vignette, 
``Extending ebnm with custom ebnm-style functions,'' available on the \pkg{ebnm} package website.
}

\subsection[The ebnm() function]{The \code{ebnm()} function}
\label{section:ebnm_function}

The \code{ebnm()} function is the main interface to the EBNM
methods. It has the following input arguments, which, apart from the first argument \code{x}, are all optional:
\begin{itemize}

\item \code{x}: The vector of observations, 
  ${\bm x} = \{x_1, \ldots, x_n\}$.

\item \code{s}: The vector of standard errors, ${\bm s} = \{s_1,
  \ldots, s_n\}$. (\code{s} may be a scalar for the homoskedastic case.)

\item \code{prior_family}: The choice of prior family $\mathcal{G}$
  (see Table~\ref{table:ebnm_prior_families}).

\item \code{mode}: For prior families that are unimodal, this argument
  specifies the location of the mode. The mode may also be estimated
  by setting \code{mode = "estimate"}.

\item \code{scale}: This is either the scale parameter (for parametric
  priors) or the grid of parameters used to approximate the
  nonparametric prior.
By default it is \code{scale = "estimate"}, which directs \pkg{ebnm} either to estimate the scale or to automatically select the grid using grid
selection strategies described in \cite{WillwerscheidDiss}.

\item \code{g_init}: An initial estimate $\hat{g}$ which can be
  used to improve the search for a maximum-likelihood estimate.

\item \code{fix_g}: A boolean value, which when TRUE causes the prior to be fixed to \code{g_init} (which must be provided) so that the posterior distributions are computed at this
  initial estimate.

\item \code{output}: A character vector indicating which 
quantities should be output.

\item \code{optmethod}: The name of the optimization method 
to use. (Currently, this option is only relevant for parametric
prior families; see Section \ref{section:ebnm_opt} below.)

\item \code{control}:
A list of control parameters to be passed to the optimization
function.

\end{itemize}

The \code{ebnm()} outputs include:
\begin{itemize}

\item \code{fitted_g}: The estimated prior $\hat{g}$.

\item \code{log_likelihood}: The log-likelihood at $\hat{g}$, which
  can be used to to compare quality of fit across different
  priors or prior families:
\begin{equation}
\textstyle \log p(x_1,\dots,x_n \mid {\bm s}, \hat{g}) = 
\sum_{i=1}^n \log \int p(x_i \mid \theta_i, s_i) \, \hat{g}(\theta_i) \, 
d\theta_i.
\end{equation}

\item \code{posterior}: Summaries of the posterior distributions
    $p(\theta_i\ |\ x_i, s_i, \hat{g})$, including posterior means, 
  posterior standard deviations and
local false sign rates \citep{Stephens_NewDeal},
\begin{equation}
\mbox{\em lfsr}(i) \colonequals
\min\big\{ p(\theta_i \le 0 \mid x_i, s_i, \hat{g}),\,
p(\theta_i \ge 0 \mid x_i, s_i, \hat{g}) \big\}.
\end{equation}
     
\item \code{posterior_sampler}: A function that can be used to
  generate random samples from the posterior distributions.

\end{itemize}
The return value is an object of class \code{"ebnm"}. Many of the S3
methods that are typically associated with model fits in \proglang{R}
also work for objects of class \code{"ebnm}", including:
\begin{itemize}

\item \code{summary()}: Gives an overview of the fitted model.

\item \code{plot()}: Produces a scatterplot comparing the observations
  $x_i$ against posterior estimates of the true means $\theta_i$ and,
  optionally, a visualization of the prior cumulative density function.

\item \code{nobs()}: Returns the number of observations $n$ used to
  fit the model.

\item \code{coef()}: Returns the posterior means from the fitted
  model, $\hat{\theta}_i \colonequals \mathbb{E}(\theta_i \mid x_i,
  s_i, \hat{g})$.

\item \code{vcov()}: Returns the posterior variances, $\mathrm{Var}(\theta_i
  \mid x_i, s_i, \hat{g})$.

\item \code{fitted()}: Returns a data frame that includes various
  posterior summary statistics for the unknowns means $\theta_i$ such
  as posterior means and variances.

\item \code{residuals()}: Returns the ``residuals'', which we define as
  the differences $x_i - \hat{\theta}_i$.

\item \code{logLik()}: Returns the log-likelihood at $\hat{g}$.

\item \code{simulate()}: Generates random draws of each $\theta_i$
  from their posterior distributions.

\item \code{quantile()}: Uses the sampler to compute posterior
  quantiles for each $\theta_i$.
  
\item \code{confint()}: Uses the sampler to compute 
  posterior ``credible intervals'' for each $\theta_i$. 
\changed{
  We define the $(1 - \alpha)$\% credible interval for $\theta_i$ as the narrowest continuous interval $[a_i, b_i]$ such that $\theta_i \in [a_i, b_i]$ with posterior probability at least $1 - \alpha$, where $\alpha \in (0,1)$. We estimate these credible intervals using Monte Carlo methods.
The proportion $1 - \alpha$ is specified by the \code{level} argument to
\code{confint()}, and is 0.95 by default.}

\item \code{predict()}: Uses the fitted prior $\hat{g}$ to compute
  posterior mean estimates $\hat{\theta}_i^{\mathrm{new}}$ for a
  different set of observations $x_i^{\mathrm{new}}$ (with standard
  deviations $s_i^{\mathrm{new}}$). This could be used, for example,
  to provide a more reliable measure of the model fit's quality by
  computing the accuracy of predictions over a test set.

\end{itemize}

We illustrate several of these methods in Section \ref{section:woba} and 
in the package vignettes.

\subsection{Details of the optimization}
\label{section:ebnm_opt}

\changed{
The \code{ebnm()} function involves two key 
computations:
\begin{enumerate}

\item {\em Estimate the prior.} 
Specifically, 
compute $\hat{g} \colonequals 
\argmax_{g \,\in\, \G} L(g)$, where
$L(g)$ denotes the marginal likelihood,
\begin{equation} 
\label{eqn:lik}
L(g) \colonequals  p({\bm x} \mid g, {\bm s}) = 
\prod_{i=1}^n \textstyle 
\int p(x_i \mid \theta_i, s_i) \, g(\theta_i) \, d\theta_i,
\end{equation}
where 
${\bm x} \colonequals (x_1, \ldots, x_n)$ and ${\bm s}
\colonequals (s_1, \ldots, s_n)$.

\item {\em Compute posterior quantities.}
\code{ebnm()} outputs various summaries 
from the
posterior distributions (means, variances, etc.) obtained using the 
estimated prior $\hat{g}$,
\begin{equation} 
p(\theta_i \mid x_i, s_i, \hat{g}) \propto 
\hat{g}(\theta_i) \, p(x_i \mid \theta_i, s_i).
\end{equation}

\end{enumerate}
The
complexity of both steps depends 
upon the choice of 
prior family $\G$ 
(see Table \ref{table:ebnm_prior_families}),
but in most cases estimating the prior is
the most difficult and computationally intensive 
step; for all but the simplest prior families,
it involves the use of numerical optimization
algorithms to compute $\hat{g}$. 
}

\begin{table}[!t]
\centering
\begin{tabular}{@{}l@{\;\;}l@{\;\;\;}c@{\;\;}c@{\;\;}c@{}} 
\code{prior\_family} & Prior & Source & Support & Sym? \\ \midrule
\multicolumn{5}{l}{\bf\!\!parametric} \\
 \code{"normal"} & $\mathcal{N}(x; \mu, \sigma^2)$ 
 & \pkg{ebnm} & $\pm$ & yes \\ 
 \code{"point_mass"} & $\delta_{\mu}(x)$ 
 & \pkg{ebnm} & & yes \\ 
\code{"point_normal"} & $\pi_0 \delta_{\mu}(x) + 
(1 - \pi_0) \mathcal{N}(x; \mu, \sigma^2)$ 
 & \pkg{ebnm} & $\pm$ & yes \\
\code{"point_laplace"} & 
$\pi_0 \delta_{\mu}(x) + (1 - \pi_0) \mathrm{Laplace}(x; \mu, a)$
& \pkg{ebnm} & $\pm$ & yes \\
\code{"point_exponential"} & 
$\pi_0 \delta_0(x) + (1 - \pi_0) \mathrm{Exp}(x; a)$ 
& \pkg{ebnm} & $+$ & no \\
\code{"horseshoe"} & $\mathrm{Horseshoe}(x; \tau)$
 & \pkg{horseshoe} & $\pm$ & yes \\[1em]
\multicolumn{5}{l}{\bf\!\!constrained nonparametric} \\
\code{"normal_scale_mixture"} & 
$\int_0^{\infty} \mathcal{N}(x; 0, \sigma^2)\,dh(\sigma^2)$
 & \pkg{ebnm} & $\pm$ & yes \\ 
\code{"unimodal_symmetric"} & $\int_0^{\infty} \mathrm{Unif}(x; -a, a)\,dh(a)$
 & \pkg{ashr} & $\pm$ & yes \\ 
\code{"unimodal"} & $\int_{-\infty}^{\infty} \mathrm{Unif}(x; 0, a)\,dh(a)$
 & \pkg{ashr} & $\pm$ & no \\
 \code{"unimodal_nonnegative"} & $\int_0^{\infty} \mathrm{Unif}(x;0,a)\,dh(a)$
 & \pkg{ashr} & $+$ & no \\[1em]
\multicolumn{5}{l}{\bf\!\!nonparametric} \\
\code{"npmle"} & $\int_{-\infty}^{\infty} \delta_t(x) \,dh(t)$
 & \pkg{ebnm} & $\pm$ & no \\ 
\code{"deconvolver"} & \cite{NarasimhanEfron} & \pkg{deconvolveR} & 
$\pm$ & no \\[1em]
\multicolumn{5}{l}{\bf\!\!other} \\
\code{"flat"} & $\mathrm{Unif}(x; -\infty, \infty)$
 & \pkg{ebnm} & $\pm$ & yes
\end{tabular}
\caption{Prior families implemented in \pkg{ebnm}. The
  ``\code{prior\_family}'' column gives the corresponding
  \code{prior\_family} argument to \code{ebnm()}. The ``Source'' column gives the name of the \proglang{R} package that
  implements the model fitting routines. ``Support''
  indicates whether the prior has support for only positive
  realizations of $\theta_i$ ($+$) or all real numbers ($\pm$). A
  ``yes'' in the ``Sym?'' column means that the prior is symmetric
  about its mode. The ``flat'' prior, which is mainly intended for use as a point of
  comparison with other prior families, is a special case with no
  parameters, and recovers the maximum-likelihood estimates
  $\hat{\theta}_i = x_i$. Note that some specialized priors such as the ``generalized binary prior''
    \citep{Yusha} are not included in this table; run \code{help(ebnm)}
    for information on such priors.}
\label{table:ebnm_prior_families}
\end{table}

\subsubsection{Parametric priors}
\label{section:ebnm_parametric}

Parametric priors available in \pkg{ebnm} include the normal,
point-normal, point-Laplace, point-exponential, and horseshoe prior families. For
normal, point-normal, and point-Laplace priors, the prior mode
can either be
estimated or fixed at zero. We  developed special implementations for
each of these prior families with the exception of the horseshoe, for which we
relied on the \pkg{horseshoe} package \citep{Horseshoe_Package}.

A closed-form solution is available only for the normal prior with
homoskedastic errors. In all other cases we use numerical methods
to search for parameter estimates maximizing the likelihood. For 
parametric prior families, this involves searching for at most three
parameters: the scale of the slab component, the mixture weight for
the spike, and, when \code{mode = "estimate"}, the mode. 

\changed{We found
that several off-the-shelf optimizers
worked well for fitting parametric priors, 
although care was needed in implementing
the underlying objective and gradient computations
to avoid numerical issues.
In particular, 
we found that the quasi-Newton method \code{nlm()} 
from the \pkg{stats} package worked very reliably in our
tests across a range of parametric prior families. Therefore,
we chose this method to be the default 
for estimating the prior in all cases except the horseshoe, which uses the \pkg{stats} function \code{optimize()} (this was the choice made 
by the authors of the \pkg{horseshoe} package).

Since other optimization methods might be
preferred in some circumstances --- say, when dealing
with large or complex data sets, or to 
refine the estimation of the prior ---
we have designed the package to allow for the use of 
other off-the-shelf optimization methods. Further, we allow the user to specify whether to use analytical gradients and Hessians, 
or whether gradients and/or Hessians are be estimated numerically 
(which is often faster, especially when the analytical calculations 
are complex). These options are controlled by the \code{optmethod} argument to \code{ebnm()}. The default for most parametric priors, \code{"nohess\_nlm"}, uses \code{nlm()} with gradients calculated analytically and Hessians estimated numerically. Alternatives include \code{"nlm"} (both gradients and Hessians are calculated analytically); \code{"nograd\_nlm"} (both gradients and Hessians are estimated numerically); \code{"lbfgsb"} and \code{"nograd\_lbfgsb"}, which use the L-BFGS-B algorithm as implemented in the \pkg{stats} function \code{optim()} 
(L-BFGS-B always estimates Hessians numerically, 
so the two options use, respectively, analytical and numerical gradients); and the trust region method from
the \pkg{trust} package \citep{trust}, which 
requires analytical gradients and 
Hessians (\code{optmethod = "trust"}). 

In our
 benchmarking experiments (Appendix~\ref{sec:optmethod}), \code{"nohess\_nlm"} 
was
either the fastest method or differed from the fastest by less than a
factor of two. All of the other \code{nlm()} methods reliably 
converged to a solution, as did the \code{trust()} method, but these other methods tended to be somewhat slower than \code{"nohess\_nlm"}. 
The L-BFGS-B methods were the least reliable; they 
occasionally failed to find a solution,
particularly in the ``null'' setting where the true prior was a point
mass at zero.
}

\subsubsection{Constrained nonparametric priors}
\label{section:ebnm_constrained_nonparametric}

\changed{
The constrained nonparametric families --- 
scale mixtures of normals and the unimodal, symmetric unimodal, and nonnegative unimodal families --- are all implemented in package \pkg{ashr}
\citep{ashr}, which uses the mix-SQP algorithm \citep{MixSQP} as its default optimization method.  Different optimization methods can again be specified via the \code{optmethod} argument to \code{ebnm()}; for details on these different optimization methods, see the documentation
in the \pkg{ashr} package. 
The only constrained nonparametric family that does not rely on \pkg{ashr}
is the family of scale mixtures of normals. For this family, we re-implemented the
\pkg{ashr} algorithm with the aim of improving efficiency. 
Our implementation improved the runtime over \pkg{ashr} 
by a full order of magnitude for 
data sets with $n \approx \mbox{1,000}$ 
(Appendix~\ref{sec:compare_with_existing_packages}).
}

\subsubsection{Nonparametric priors}
\label{section:ebnm_unconstrained_nonparametric}

\changed{
The NPMLE can, in principle, be computed using \pkg{ashr}, but this computation is cumbersome since \pkg{ashr} requires the user to specify the grid of point
masses in advance. Further, we have found that, as with scale mixtures of normals, 
\pkg{ashr} can be slow for large data sets. The \pkg{REBayes} package
\citep{KoenkerGu} was developed specifically for the NPMLE, and is
typically very fast, but it relies on the commercial interior-point
solver \proglang{MOSEK} \citep{mosek}. Therefore, in order to 
provide a fully open-source toolkit that does not require installation
of commercial software, we re-implemented estimation of the
NPMLE in \pkg{ebnm} using the open-source package \pkg{mixsqp}
\citep{MixSQP}. As with the constrained nonparametric prior families, 
\code{optmethod = "mixsqp"} is the default
setting. If desired, however, the \pkg{REBayes} algorithm can be used by setting \code{optmethod = "REBayes"}.
In our tests, \pkg{mixsqp} was typically faster than \pkg{REBayes} for smaller
$K$ (the number of mixture components in the prior; see eq.~\ref{eq:g_npmle}), 
whereas \pkg{REBayes} was often faster than \pkg{mixsqp} when $K$ 
approached or exceeded 80 
(see Appendix \ref{sec:compare_with_existing_packages}).}

\section{Numerical comparisons of prior families}
\label{section:ebnm_simdata}

To test our implementations and to compare prior families, we simulated  data sets from three
different data-generating distributions:
\begin{enumerate}

\item {\bf Normal}. In this simplest case, we simulated from a
  normal prior, $\theta_i \sim \mathcal{N}(0,2^2)$.

\item {\bf Point-${\bm t}$}. In this second, more challenging scenario,
  the prior was both sparse and heavy-tailed, yet still symmetric:
  $\theta_i \sim 0.8 \delta_0 + 0.2 t_5(0, 1.5)$, where
  $t_{\nu}(\mu, \sigma)$ denotes the Student-$t$ distribution with
  location $\mu$, scale $\sigma$, and $\nu$ degrees of freedom.

\item {\bf Asymmetric Tophat}. In the third simulation scenario, we
  simulated data with uniformly-distributed means, $\theta_i \sim
  \mathrm{Unif}(-5, 10)$. Although this scenario is perhaps less realistic than the
  other simulations, it yields data sets that are best modeled with nonparametric or constrained nonparametric
  priors.

\end{enumerate}
In all simulations, we generated the observations as $x_i \sim
\mathcal{N}(\theta_i, 1)$.

\begin{figure}[!t]
\centering
\includegraphics[width=0.975\textwidth]{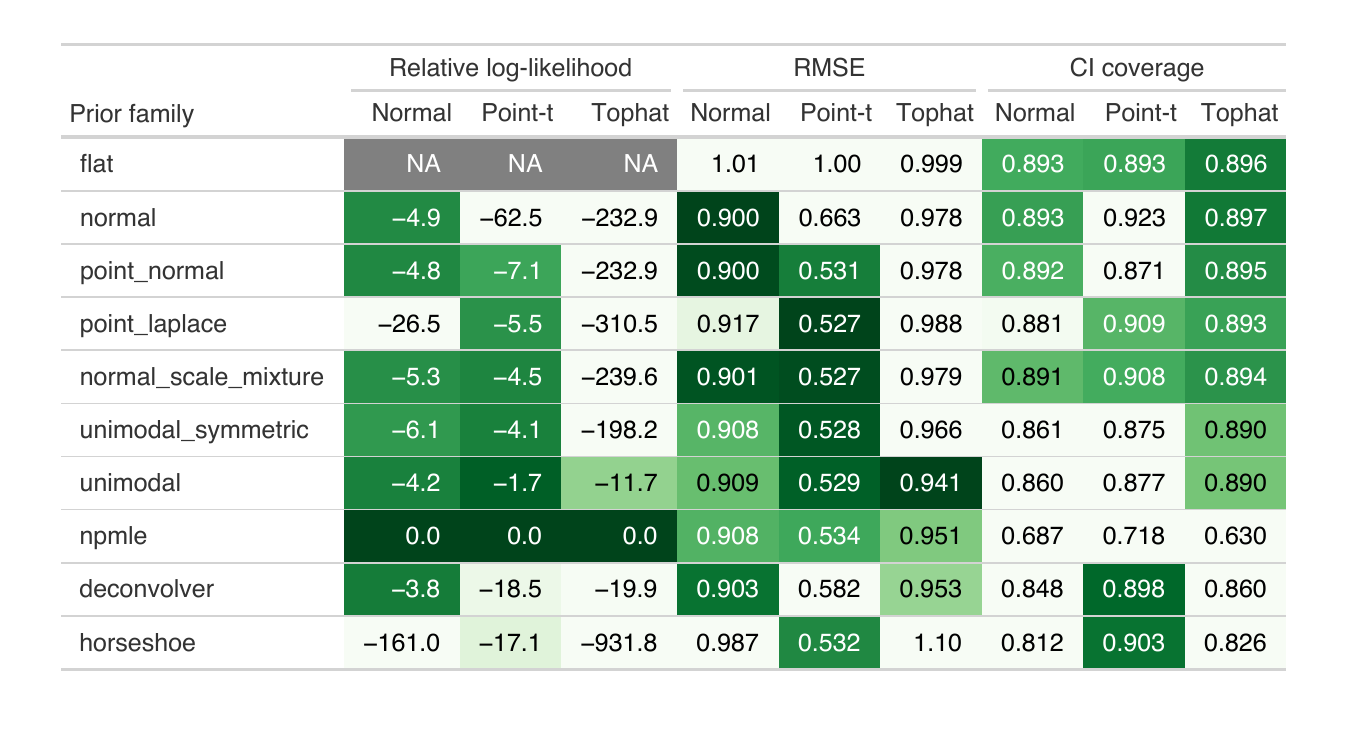}
\caption{Results of fitting EBNM models with different prior families
  to ``Normal'', ``Point-$t$'' and ``Asymmetric tophat'' data sets.
  For each prior family, the table gives the following quantities
  averaged across EBNM analyses of 20 data sets with $n =
  \mbox{1,000}$ observed means, with ``better'' results highlighted using darker shades of green: the log-likelihood at $\hat{g}$
  relative to the log-likelihood attained by the NPMLE (higher log-likelihoods are
  better); the root mean-squared error (lower RMSEs are better); and the proportion of 90\% posterior credible intervals containing the true mean (CI coverage values
  closer to 0.9 are better).}
\label{fig:simres}
\end{figure}

Figure~\ref{fig:simres} summarizes results from running EBNM
analyses on 10 data sets in each of the three simulation scenarios,
with $n = \mbox{1,000}$ observations in each data set. We used the
following three measures to evaluate the EBNM model fits:
\renewcommand{\labelenumi}{\alph{enumi}.}
\begin{enumerate}

\item The log-likelihood, which, for ease of interpretation, is shown
  relative to the log-likelihood attained at the NPMLE estimate. (In
  theory, the NPMLE estimate should always give the highest
  likelihood because the family of all distributions includes all other prior families as proper subsets.) These log-likelihoods were obtained by calling
  \code{logLik()} on the \code{ebnm()} return value.

\item The root mean-squared error, $\mathrm{RMSE}
  \colonequals \sqrt{\sum_{i = 1}^n(\hat{\theta}_i - \theta_i)^2/n}$,
  where $\hat{\theta}_i$ denotes the posterior mean estimate,
  $\hat{\theta}_i \colonequals \mathbb{E}(\theta_i \mid x_i, s_i,
  \hat{g})$. These estimates were obtained by calling \code{coef()} on
  the \code{ebnm()} return value.

\item The proportion of true means $\theta_i$ that are contained
  within the 90\% posterior credible intervals, which were obtained by calling \code{confint()} on the \code{ebnm()} return
  value. (For our method for computing credible intervals, see Section \ref{section:ebnm_function} above.)

\end{enumerate}

As expected, the model fit returned by \code{ebnm()} with
\code{prior\_family = "npmle"} always attained the largest
log-likelihood.  More generally, log-likelihoods were largely (though not exactly) aligned
with the orderings implied by nestings of prior families, such as, for example,
\begin{equation} \label{eqn:nesting}
\mathcal{G}_{\text{norm0}} \subset
\mathcal{G}_{\text{pn}} \subset
\mathcal{G}_{\text{smn}} \subset
\mathcal{G}_{\text{symm-u}} \subset
\mathcal{G}_{\text{npmle}}.
\end{equation}
Prior families that were a poor match with the distribution used to
simulate the data typically had worse log-likelihoods.

The RMSE evaluates the quality of the posterior estimates
$\hat{\theta}_i$ generated by an EBNM analysis. Reassuringly, nearly all prior families improved
upon the maximum-likelihood estimates $\hat{\theta}_i = x_i$ returned by the ``flat'' prior, which we included as a baseline. However, the improvement was
sometimes small, particularly when the prior family was a poor match
with the true distribution (e.g., symmetric prior families in the
asymmetric tophat scenario).  
In general, higher log-likelihoods were indicative of better
accuracy in estimates of $\theta_i$. Exceptions are suggestive of overfitting; for example, the RMSE for the NPMLE was typically worse than for prior families that better matched the true distribution.

The ``CI coverage'' measures how well posterior credible intervals are
calibrated.  A known limitation of empirical Bayes methods is that
they often underestimate uncertainty in the posteriors, since
uncertainty in the estimate of $g$ is not taken into account \citep[see, for example,][]{ignatiadis2022confidence}.
Indeed, the credible intervals tended to be too small (i.e., less than 90\%) for most prior
families and simulation scenarios (Figure~\ref{fig:simres}). Still, the intervals were usually not far off the target coverage of 90\%,
showing a surprising robustness to modeling
assumptions.
The lone exception was the NPMLE, which tended to have much
poorer coverage because, as noted in Section \ref{section:nonparametric}, it results in a discrete
prior that can greatly underestimate uncertainty.

\begin{figure}[!t]
\centering
\includegraphics[width=0.85\textwidth]{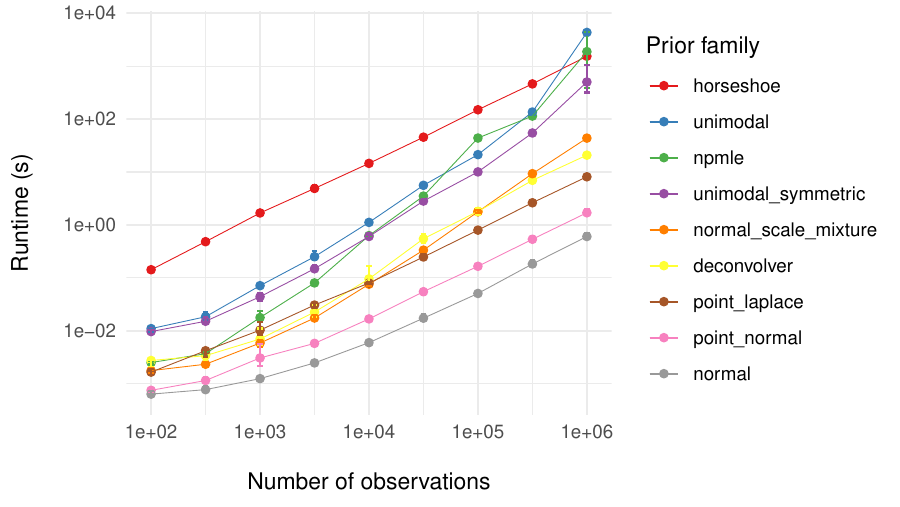}
\caption{Runtimes for fitting EBNM models with different prior 
  families to data sets ranging in size from $n = 100$ to $n =
  \mbox{1,000,000}$. For each combination of sample size ($n$) and
  prior family, an EBNM model was fit to 20 data sets simulated using
  the Point-$t$ distribution. Each point in the plot gives the average
  runtime over all 20 simulations; the error bars depict  10\%
  and 90\% quantiles.}
\label{fig:timecomps}
\end{figure}

Finally, to assess the ability of our implementation to handle large
data sets, we recorded runtimes for simulated (point-$t$) data sets
ranging in size from $n = 200$ to $n = \mbox{1,000,000}$. These
analyses were performed in \proglang{R} 4.2.2 on a desktop running Windows 11 Pro 
with an Intel Core i9-13900KF multicore processor and 32 GB of memory. Results are summarized in
Figure~\ref{fig:timecomps}. As expected, the less flexible priors with
the fewest parameters tended to also be the fastest, whereas the most
complex methods (e.g., unimodal prior, NPMLE) were slower than the
fastest methods by multiple orders of magnitude.  Most importantly, all
prior families implemented in \pkg{ebnm} scaled well to large data
sets; the computational effort grew linearly or close to linearly in
$n$.

\section[An analysis of weighted on-base averages with ebnm]{An analysis of weighted on-base averages with \pkg{ebnm}}
\label{section:woba}

\changed{In this section, we illustrate the key features of \pkg{ebnm}
  in an analysis of baseball statistics. See the package
  vignette for an expanded version of this example.}

\subsection{The ``wOBA'' data set}

\changed{
We begin by loading and inspecting the \code{wOBA} data set, which
consists of wOBAs (``weighted on-base averages'') and standard errors
for the 2022 MLB regular season:}
\begin{CodeChunk}
\begin{CodeInput}
R> library("ebnm")
R> data("wOBA")
R> nrow(wOBA)
\end{CodeInput}
\begin{CodeOutput}
[1] 688
\end{CodeOutput}
\begin{CodeInput}
R> head(wOBA)
\end{CodeInput}
\begin{CodeOutput}
  FanGraphsID           Name Team  PA     x     s
1       19952     Khalil Lee  NYM   2 1.036 0.733
2       16953 Chadwick Tromp  ATL   4 0.852 0.258
3       19608     Otto Lopez  TOR  10 0.599 0.162
4       24770   James Outman  LAD  16 0.584 0.151
5        8090 Matt Carpenter  NYY 154 0.472 0.054
6       15640    Aaron Judge  NYY 696 0.458 0.024
\end{CodeOutput}
\end{CodeChunk}
\changed{Column ``x'' contains observed wOBAs, which we interpret
as estimates of a player's {\em hitting ability}.
Column ``s'' gives standard errors. See Appendix \ref{sec:woba} for
background on the wOBA statistic and details on how standard errors
were calculated.}

\changed{ Most players finished the season with a wOBA between .200
  and .400.
A few had very high wOBAs ($>$.500), while others had wOBAs at or near
zero. A casual inspection of the data suggests that players with these extreme wOBAs were simply lucky (or unlucky). For
example, the 4 players with the highest wOBAs (included in the code output above) each had fewer than 20 plate appearances. 
(The number of plate appearances, or PAs, is the sample size over which wOBA is measured for each hitter, so smaller numbers of PAs are generally associated with larger standard errors.)
}

\changed{In contrast, Aaron Judge's production --- which included a
record-breaking number of home runs --- appears to be ``real,'' since
it was sustained over nearly 700 PAs.
Other cases are more ambiguous: how, for example, are we to assess
Matt Carpenter, who had several exceptional seasons between 2013 and
2018 but whose output steeply declined in 2019--2021 before his
surprising ``comeback'' in 2022? An empirical Bayes analysis can help
to answer this and other questions.}

\subsection{The ``ebnm'' function}

\changed{Function \code{ebnm()} is the main interface for fitting the
  empirical Bayes normal means model
  (\ref{eqn:nm_problem}--\ref{eqn:nm_prior}); it is a ``Swiss army
  knife'' that allows for various choices of prior family
  $\mathcal{G}$ as well as multiple options for fitting and
  tuning models. For example, we can fit a normal means model with $\mathcal{G}$ taken to be the family of normal
  distributions as follows:}
\begin{CodeChunk}
\begin{CodeInput}
R> x <- wOBA$x
R> s <- wOBA$s
R> names(x) <- wOBA$Name
R> names(s) <- wOBA$Name
R> fit_normal <- ebnm(x, s, prior_family = "normal", mode = "estimate")
\end{CodeInput}
\end{CodeChunk}
\changed{
(The default behavior is to fix the prior mode at zero.
Since we certainly do not expect the distribution of true hitting ability to be centered at zero, we set \code{mode = "estimate"}.)}

\changed{The \pkg{ebnm} package has a second model-fitting interface
  in which each prior family gets its own function:}
\begin{CodeChunk}
\begin{CodeInput}
R> fit_normal <- ebnm_normal(x, s, mode = "estimate")
\end{CodeInput}
\end{CodeChunk}

\begin{figure}[!t]
\centering
\includegraphics[width=0.7\textwidth]{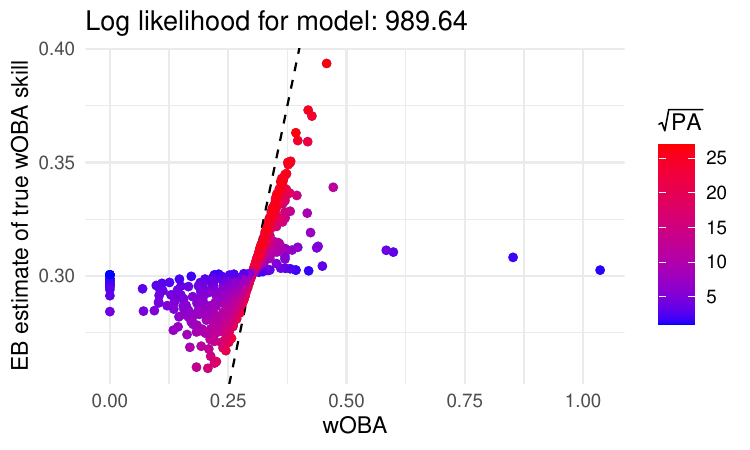}
\caption{Initial wOBA estimates (``observations'') vs. posterior mean
  wOBA estimates, in which the posterior estimates were obtained by
  fitting a prior from the family of normal distributions. The color
  of the points is varied by the number of plate appearances. The
  dashed line shows the diagonal ($x = y$) line.}
\label{fig:wOBA_normal_custom}
\end{figure}

\changed{Textual and graphical overviews of results can be obtained
  using the \code{summary()} and \code{plot()} methods.
The \code{plot()} method returns a \code{"ggplot"} object \citep{ggplot2},
so that the plot can be conveniently customized using \pkg{ggplot2}. For
example, we can vary the color of points by the number of plate
appearances:}
\begin{CodeChunk}
\begin{CodeInput}
R> plot(fit_normal) +
+    geom_point(aes(color = sqrt(wOBA$PA))) +
+    labs(x = "wOBA", y = "EB estimate of true wOBA skill", 
+       color = expression(sqrt(PA))) +
+    scale_color_gradient(low = "blue", high = "red")
\end{CodeInput}
\end{CodeChunk}
\changed{The resulting plot, shown in
  Figure~\ref{fig:wOBA_normal_custom}, compares the initial wOBA
  estimates --- that is, the first input in the \code{ebnm()} call ---
  against the posterior estimates returned by \code{ebnm()}.
The plot tells us that wOBAs associated with fewer plate appearances
(blue points) were shrunk toward the league average (near .300) much
more strongly than wOBAs for hitters with many plate appearances (red
points).  
}

\changed{Let us revisit the first 6 hitters in the data set to see
  what the EBNM model suggests about their true hitting ability. The
  \code{fitted()} method returns a posterior summary for each hitter
  (by default, the posterior mean and standard deviation):}
\begin{CodeChunk}
\begin{CodeInput}    
> print(head(fitted(fit_normal)), digits = 3)
\end{CodeInput}
\begin{CodeOutput}
                mean     sd
Khalil Lee     0.303 0.0287
Chadwick Tromp 0.308 0.0286
Otto Lopez     0.310 0.0283
James Outman   0.311 0.0282
Matt Carpenter 0.339 0.0254
Aaron Judge    0.394 0.0184
\end{CodeOutput}
\end{CodeChunk}
\changed{Estimates for the first four ballplayers are shrunk strongly
toward the league average, reflecting the fact that these players had
very few plate appearances.
Carpenter had many more plate appearances (154) than these other
four players, but according to this model we should remain skeptical about
his strong performance; after factoring in the prior, we judge his
``true'' talent to be much closer to the league average,
downgrading an observed wOBA of .472 to the posterior mean
estimate of .339.}

\subsection{Comparing different priors}

\changed{Judge's ``true'' talent is also estimated to be much lower (.394) than
his observed wOBA (.458) despite sustaining this high level of
production over a full season (696 PAs). For this reason, one might
ask whether a prior that is more flexible than the normal prior ---
that is, a prior that can better adapt to ``outliers'' like Judge ---
might produce a different result. The \pkg{ebnm} package is very well
suited to answering this question. For example, to obtain results
using the family of all unimodal priors rather than a normal prior, we need only
update the argument to \code{prior\_family}:}
\begin{CodeChunk}
\begin{CodeInput}
R> fit_unimodal <- ebnm(x, s, prior_family = "unimodal", mode = "estimate")
\end{CodeInput}
\end{CodeChunk}
\changed{
Using this prior, estimates for
players with many plate appearances and outlying performances (very high or very low wOBAs) are not adjusted quite so
strongly toward the league average. 
Judge's estimated ``true'' talent, for example, remains much closer to his observed wOBA:}
\begin{CodeChunk}
\begin{CodeInput}
R> dat <- cbind(wOBA[, c("PA","x")],
+               fitted(fit_normal),
+               fitted(fit_unimodal))
R> names(dat) <- c("PA", "x", "mean_n", "sd_n", "mean_u", "sd_u")
R> print(head(dat), digits = 3)
\end{CodeInput}
\begin{CodeOutput}
                PA     x mean_n   sd_n mean_u   sd_u
Khalil Lee       2 1.036  0.303 0.0287  0.302 0.0277
Chadwick Tromp   4 0.852  0.308 0.0286  0.307 0.0306
Otto Lopez      10 0.599  0.310 0.0283  0.310 0.0315
James Outman    16 0.584  0.311 0.0282  0.311 0.0318
Matt Carpenter 154 0.472  0.339 0.0254  0.355 0.0430
Aaron Judge    696 0.458  0.394 0.0184  0.439 0.0155
\end{CodeOutput}
\end{CodeChunk}
\changed{Carpenter's estimated ``true'' talent is also higher, but is still adjusted much more than Judge's in light
of Carpenter's smaller sample size. Interestingly,
the
unimodal prior also assigns greater uncertainty (the ``sd\_u'' column)
to Carpenter's estimate than does the normal prior. 
}

\subsection{Reanalysis using a nonparametric prior}

\changed{
An alternative to prior families that make
specific assumptions about the data is to use the prior family that
contains \textit{all} distributions $\mathcal{G}_{\mathrm{npmle}}$,
which is in a sense ``assumption free'' (see Section
\ref{section:nonparametric} for background). Note that
although nonparametric priors require specialized computational
techniques, switching to a nonparametric prior is seamless in
\pkg{ebnm}, as these implementation details are hidden. Similar to
above, we need only make a single change to the \code{prior\_family}
argument:}
\begin{CodeChunk}
\begin{CodeInput}
R> fit_npmle <- ebnm(x, s, prior_family = "npmle")
\end{CodeInput}
\end{CodeChunk}
\changed{(Note that because the family $\mathcal{G}_{\mathrm{npmle}}$
  is not unimodal, the \code{mode = "estimate"} option is not relevant
  here.)}

\begin{figure}[!t]
\centering
\includegraphics[width=0.75\textwidth]{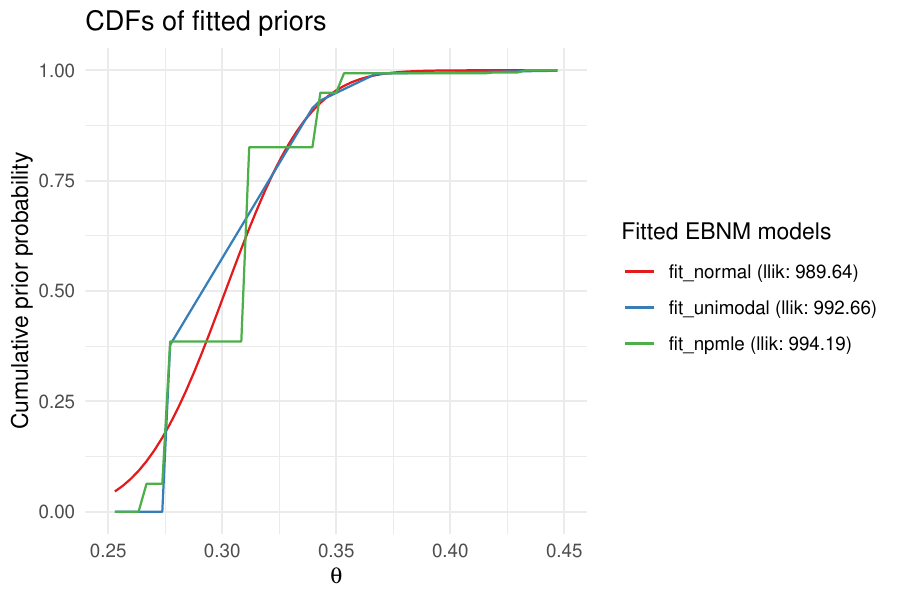}
\caption{The CDFs for priors fitted to the 2022 MLB wOBA data: the
  normal prior (\code{prior\_family = "normal"}), the unimodal prior
  (\code{prior\_family = "unimodal"}), and the NPMLE
  (\code{prior\_family = "npmle"}).}
\label{fig:wOBA_npmle_cdf}
\end{figure}

\begin{figure}[!t]
\centering
\includegraphics[width=0.75\textwidth]{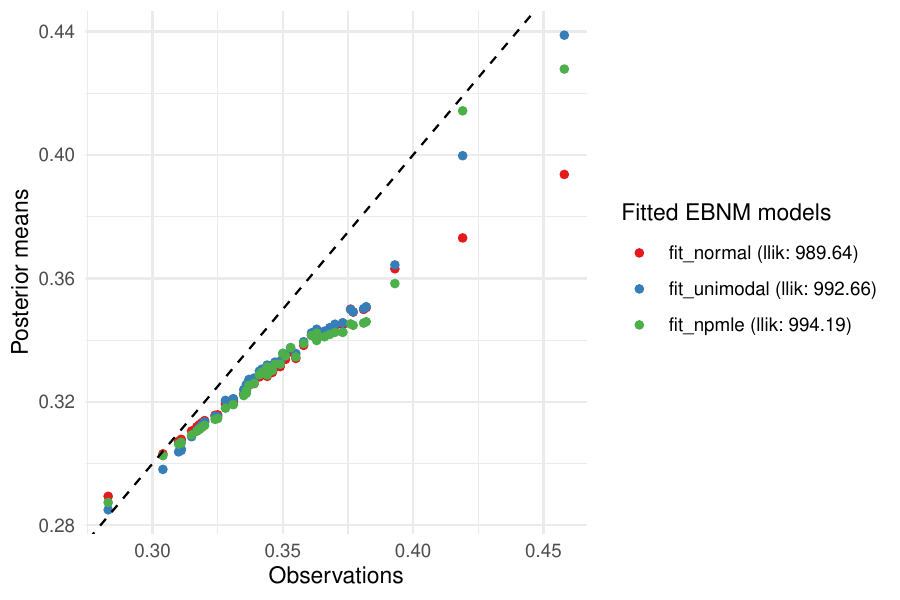}
\caption{Initial wOBA estimates (``Observations'') vs. posterior mean
  wOBA estimates for priors fitted to the 2022 MLB wOBA data: the
  normal prior (\code{prior\_family = "normal"}), the unimodal prior
  (\code{prior\_family = "unimodal"}), and the NPMLE
  (\code{prior\_family = "npmle"}). Results are shown only for the top
  50 ballplayers by number of plate appearances.}
\label{fig:wOBA_npmle_pm}
\end{figure}

\changed{We visually compare the three fits obtained so far using the \code{plot()} method. 
We use the \code{subset} argument to focus on results for
Judge and other players with a large number of plate appearances, and 
we include argument 
\code{incl\_cdf =
   TRUE} to also show the cumulative distribution functions (CDFs) of the
   fitted priors $\hat{g}$:}
\begin{CodeChunk}
\begin{CodeInput}
top50 <- order(wOBA$PA, decreasing = TRUE)
top50 <- top50[1:50]
plot(fit_normal, fit_unimodal, fit_npmle, incl_cdf = TRUE, subset = top50)
\end{CodeInput}
\end{CodeChunk}
\changed{The plots generated by this call are shown in Figures
\ref{fig:wOBA_npmle_cdf} and \ref{fig:wOBA_npmle_pm}.  Estimates largely agree, differing primarily at the tails (e.g., Judge), where both the
unimodal prior family and the NPMLE are sufficiently flexible to avoid
the strong shrinkage behavior of the normal prior family.}

\changed{Fits can be compared quantitatively using the \code{logLik()} method,
which, in addition to the log likelihood for each model, usefully
reports the number of free parameters or ``degrees of freedom'' (note that Wilks' theorem does not apply to these nonparametric comparisons):}
\begin{CodeChunk}
\begin{CodeInput}
R> logLik(fit_unimodal)
\end{CodeInput}
\begin{CodeOutput}
'log Lik.' 992.6578 (df=40)
\end{CodeOutput}
\begin{CodeInput}
R> logLik(fit_npmle)
\end{CodeInput}
\begin{CodeOutput}
'log Lik.' 994.193 (df=94)
\end{CodeOutput}
\end{CodeChunk}
\changed{A nonparametric prior $\mathcal{G}$ is approximated by $K$ mixture
components on a fixed grid, with mixture proportions to be
estimated (see Section~\ref{section:constrained_nonpar}). We can infer
from the above output that the family of unimodal priors has been approximated by a family
of mixtures over $K = 41$ fixed components, while $\mathcal{G}_\text{npmle}$ has been
approximated as a family of mixtures over a grid of $K = 95$ point
masses spanning the range of the data. (The number of degrees of
freedom is one fewer than $K$ because the mixture proportions must
always sum to 1, which removes one degree of freedom from the
estimation of ${\bm\pi}$.)}

\changed{One potential issue with the NPMLE is that, since it is discrete (as
Figure~\ref{fig:wOBA_npmle_cdf} makes apparent),
observations are variously shrunk toward one of the support points,
which can result in poor interval estimates. For illustration, we
calculate 80\% posterior credible intervals (since credible intervals are obtained using Monte Carlo methods, we set a seed for reproducibility):}
\begin{CodeChunk}
\begin{CodeInput}
R> fit_npmle <- ebnm_add_sampler(fit_npmle)
R> set.seed(123)
R> print(head(confint(fit_npmle, level = 0.8)), digits = 3)
\end{CodeInput}
\begin{CodeOutput}
               CI.lower CI.upper
Khalil Lee        0.265    0.309
Chadwick Tromp    0.276    0.342
Otto Lopez        0.276    0.342
James Outman      0.276    0.342
Matt Carpenter    0.309    0.419
Aaron Judge       0.430    0.430
\end{CodeOutput}
\end{CodeChunk}
\changed{Each credible interval endpoint is constrained to lie at one
of the support points of the NPMLE $\hat{g}$.
Note in particular that
the NPMLE yields a degenerate interval estimate for Judge.}

\changed{To address this and other issues, the \pkg{deconvolveR} package
\citep{NarasimhanEfron}
uses a penalized likelihood that encourages ``smooth'' priors
$\hat{g}$; that is, priors $\hat{g}$ for which few of the mixture
proportions are zero:}
\begin{CodeChunk}
\begin{CodeInput}
R> fit_deconv <- ebnm_deconvolver(x / s, output = ebnm_output_all())
\end{CodeInput}
\end{CodeChunk}
\changed{
Note however that since package \pkg{deconvolveR}
fits a model to $z$-scores rather than observations and associated
standard errors, the ``true'' means $\theta$ being estimated are
$z$-scores rather than raw wOBA skill. While this may be  reasonable in many settings, it does
not seem appropriate for the wOBA data:}
\begin{CodeChunk}
\begin{CodeInput}
R> set.seed(123)
R> print(head(confint(fit_deconv, level = 0.8) * s), digits = 3)
\end{CodeInput}
\begin{CodeOutput}
               CI.lower CI.upper
Khalil Lee        0.000    1.600
Chadwick Tromp    0.563    1.127
Otto Lopez        0.442    0.796
James Outman      0.412    0.742
Matt Carpenter    0.413    0.531
Aaron Judge       0.406    0.459
\end{CodeOutput}
\end{CodeChunk}
\changed{
These interval estimates do not match our basic intuitions;
for example, a wOBA over .600 has never been sustained over a
full season.}

\section[Building on ebnm for new matrix factorization methods]{Building on \pkg{ebnm} for new matrix factorization methods}
\label{section:ebnm_snmf}

\changed{As mentioned above in the introduction, 
the EBNM model underlies other well-studied statistical problems,
and so there is the potential for \pkg{ebnm} to aid in the 
development of other software tools.
One such example is 
{\em matrix factorization}: as \cite{WangStephens} showed, fitting  an empirical Bayes matrix factorization (EBMF) model
can be reduced to solving a sequence of EBNM problems (typically very many of them).
Therefore, the aspects that we have emphasized in developing \pkg{ebnm} --- 
the unified interface, 
the variety of prior families and fitting options, 
and the speed and robustness of the numerical optimization 
 --- have greatly facilitated the 
creation of a flexible software framework for EBMF in the \proglang{R} 
package \pkg{flashier} \citep{flashier}, available on CRAN and GitHub 
(\url{https://github.com/willwerscheid/flashier/})}

\changed{
In matrix factorization, we attempt to approximate a data matrix ${\bf X}$ by
a low-rank matrix product, ${\bf X} \approx {\bf L} {\bf F}^\top$.
The EBMF approach introduces priors on the low-rank matrices ${\bf L}$ and ${\bf F}$:
\begin{equation} 
\label{eqn:ebmf}
\begin{aligned}
\mathbf{X} &= \mathbf{L} \mathbf{F}^\top + \mathbf{E} \\
e_{ij} &\sim \mathcal{N}(0, \sigma^2) \\
\ell_{ik} &\sim g_\ell^{(k)} \in \mathcal{G}_{\ell} \\
f_{jk} &\sim g_f^{(k)} \in \mathcal{G}_f, 
\end{aligned}
\end{equation}
where 
${\bf X}$, ${\bf L}$, ${\bf F}$, and ${\bf E}$ are, respectively, matrices of
dimension $n \times p$, $n \times K$, $p \times K$, and $n \times p$ storing
real-valued elements $x_{ij}$, $l_{ik}$, $f_{jk}$, and $e_{ij}$,
and $\mathcal{G}_{\ell}, \mathcal{G}_f$ are specified prior families. 
In brief, each iteration of the EBMF model-fitting algorithm involves solving an EBNM
problem separately for each column of ${\bf L}$ (using the prior family $\mathcal{G}_{\ell}$) and each column of
${\bf F}$ (using the prior family $\mathcal{G}_f$). 
The solutions to these EBNM problems yield fitted priors $\hat{g}_{\ell}^{(k)}, \hat{g}_f^{(k)}$
and posterior estimates of $\ell_{ik}$ and $f_{jk}$. See \cite{WangStephens} for details.}

\changed{The EBMF framework is highly flexible in that different choices of prior
families $\mathcal{G}_{\ell}$ and $\mathcal{G}_f$ can give very different factorizations. 
For example, the use of normal priors yields factorizations similar to
the truncated singular value decomposition (SVD) \citep{Nakajima}. 
The use of {\em sparse priors} (e.g., the point-normal prior family) can yield {\em sparse matrix factorizations}, which in many settings are more interpretable than an SVD \citep{Engelhardt, SSVD, Witten}. By choosing priors with nonnegative support (e.g., the point-exponential family), one can obtain {\em nonnegative factorizations} \citep{LeeSeung}. More novel 
combinations are also possible: 
for example, one can obtain a {\em semi-nonnegative matrix factorization} \citep{DingJordan,
wang2019three, he2020sn} by choosing a prior family with nonnegative support for $\mathcal{G}_{\ell}$ and a prior family without constraints for $\mathcal{G}_f$; and \cite{Yusha} proposed the family of ``generalized binary'' priors to encourage binary-valued $l_{ik}$.}

\changed{By building on the fast and reliable methods in \pkg{ebnm}, the 
\pkg{flashier} package makes it straightforward to obtain any of these kinds of matrix factorization (and many more).
For example, a sparse factorization can be obtained by calling the 
\pkg{flashier} function \code{flash()} with argument \code{ebnm\_fn = ebnm\_point\_normal}, which 
specifies point-normal distributions for all priors $g_{\ell}^{(k)}$ and $g_f^{(k)}$. To obtain a sparse, semi-nonnegative factorization, one need only update the argument as \code{ebnm\_fn = c(ebnm\_point\_exponential, ebnm\_point\_normal)}, 
which specifies point-normal priors for all $g_f^{(k)}$ and 
point-exponential priors for all $g_{\ell}^{(k)}$.
In general, any of the 
prior families discussed above can be used (see Table \ref{table:ebnm_prior_families}), and if some other option is desired, it is not difficult to implement a new
``ebnm-style'' function (see the \pkg{ebnm} package vignette for details).
}

\changed{
We provide a detailed illustration of these ideas 
in the \pkg{flashier} package vignette, 
``Introduction to flashier,'' available on the package's website (\url{https://willwerscheid.github.io/flashier/}). 
}

\section{Summary}
\label{sec:summary}

The \pkg{ebnm} package provides a comprehensive toolkit for solving the 
empirical Bayes normal means (EBNM) problem under a variety of prior assumptions. 
In many situations --- as in our analysis of baseball 
statistics in Section~\ref{section:woba} --- 
the ``best'' choice of prior family is not known in advance. 
The \pkg{ebnm} package is especially well-suited
to handling such situations by providing a large set of prior families to 
choose from (Table~\ref{table:ebnm_prior_families}), and an interface 
that allows for convenient comparison of different prior families. 
When deciding which prior family to proceed with, 
our general recommendation is to weigh prior assumptions about the data 
against empirical measures of fit.
\changed{The best prior will very often depend on the context, and
for this reason we have designed \pkg{ebnm} to be easily extensible so that 
researchers are not limited by the existing options. 
Our ultimate hope is that experts in other research areas will consider contributing to our package
and help expand the use of EBNM methods to
other domains.}

\bibliography{ebnm}

\appendix

\section{Supplementary benchmarking results}
\label{section:ebnm_benchmarks}

\subsection{Optimization methods for parametric families}
\label{sec:optmethod}

We first compare the performance of six optimization methods, all of which are implemented in \pkg{ebnm} via parameter \code{optmethod}. In each case, parameters are transformed so that the optimization problem is unconstrained (specifically, a log transformation is used for scale parameters, which are constrained to be nonnegative, while a logit transformation is used for mixture proportions, which are constrained to lie between zero and one). 

Three choices of \code{optmethod} call into function \code{nlm()}, a Newton-type algorithm included in the base \pkg{stats} package. Gradient and Hessian functions can be provided; if they are not, \code{nlm()} estimates them numerically. Option \code{optmethod = "nlm"} provides both the gradient and Hessian functions; \code{optmethod = "nohess\_nlm"} provides the gradient but not the Hessian; \code{optmethod = "nograd\_nlm"} provides neither. Options \code{optmethod = "lbfgsb"} and \code{optmethod = "nograd\_lbfgsb"} call into function \code{optim()}, also in the \pkg{stats} package, with argument \code{method = "L-BFGS-B"}. The former provides the gradient function; the second does not. By definition, L-BFGS-B does not accept a Hessian. Finally, \code{optmethod = "trust"} calls into function \code{trust()}, a trust-region algorithm implemented in the \pkg{trust} package \citep{trust}. Since \code{trust()} requires both a gradient and Hessian function, there is only one corresponding \code{optmethod}. In sum, then, there are two methods that use both gradients and Hessians (\code{"nlm"} and \code{"trust"}); two that use only gradients (\code{"nohess\_nlm"} and \code{"lbfgsb"}); and two that estimate all derivatives numerically (\code{"nograd\_nlm"} and \code{"nograd\_lbfgsb"}).

For both \code{ebnm\_point\_normal()} and \code{ebnm\_point\_laplace()}, we ran tests for $2 \times 3 \times 2 \times 3 = 36$ scenarios: 

\begin{itemize}
    \item The mode is either fixed at zero via argument \code{mode = 0} or estimated (via \code{mode = "estimate"}).
    \item The data-generating prior distribution $g$ is: i) a true member of the prior family, $\pi_0 \delta_\mu + (1 - \pi_0) N(\mu, a^2)$ or $\pi_0 \delta_\mu + (1 - \pi_0) \text{Laplace}(\mu, a)$, with $\pi_0 \sim \text{Beta}(10, 2)$, $a \sim \text{Gamma}(4, 1)$, and either $\mu = 0$ or $\mu \sim \text{Unif}(-10, 10)$; ii) the null distribution $\delta_0$; or iii) a distribution not in the prior family, so that $\mathcal{G}$ is misspecified. When \code{mode = 0}, the misspecified prior is a point-normal prior as above but with $\mu \sim \text{Unif}(-10, 10)$; when \code{mode = "estimate"}, the misspecified prior is the point-$t_5$ distribution $\pi_0 \delta_0 + (1 - \pi_0) t_5(0, a)$, with $\pi_0$ and $a$ distributed as above.
    \item The $N(0, s_i^2)$ noise added to the true means $\theta_i \sim g$ is either homoskedastic, with $s_i = 1$ for all $i$, or heteroskedastic, with $s_i^2 - 1 \sim \text{Exp}(1)$.
    \item The number of observations $n$ is 1000, 10000, or 100000.
\end{itemize}

\begin{figure}
    \centering
    \includegraphics[width=.9\textwidth]{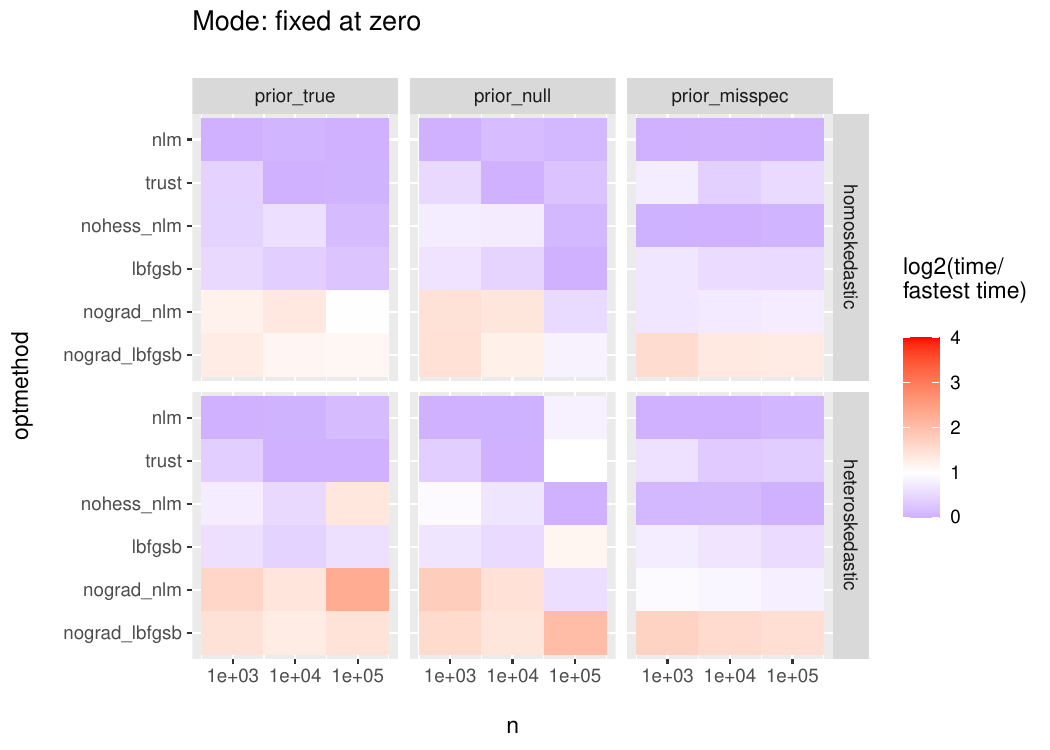}
    \includegraphics[width=.9\textwidth]{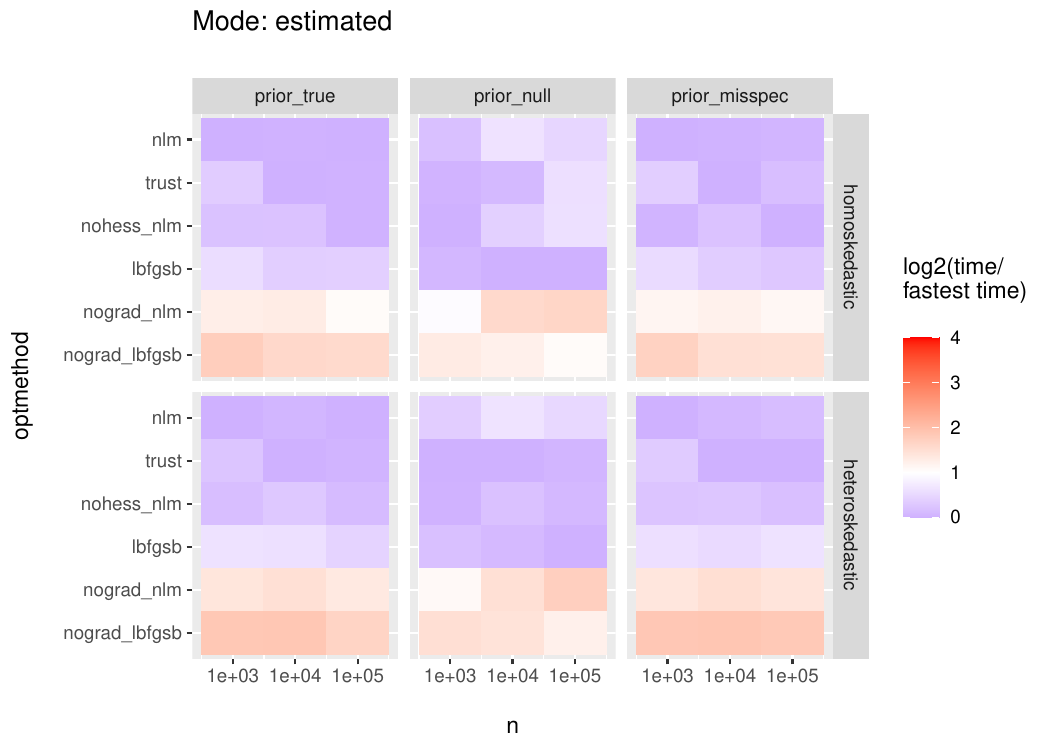}
    \caption{Timing comparisons for \code{ebnm\_point\_normal()}. Tests in the top figure fix the mode at zero (by setting argument \code{mode = 0}) while those in the bottom estimate the mode (by setting \code{mode = "estimate"}). Different columns correspond to different data-generating priors: a true member of the point-normal prior family; the null distribution $\delta_0$; or a distribution from outside the prior family. Different rows correspond to different noise models, with the noise added to the ``true'' observations either homoskedastic (with $s_i = 1$ for all $i$) or heteroskedastic (with $s_i^2 - 1 \sim \text{Exp}(1)$). Plotted is the time as a multiple of the fastest time for a given combination of mode, prior, noise, and value of $n$. Shades of blue indicate optimization methods that are within a factor of 2 of the fastest method for that mode, prior, noise, and value of $n$, while shades of red indicate slower methods.}
    \label{fig:ebnm_pn}
\end{figure}

\begin{figure}
    \centering
    \includegraphics[width=.9\textwidth]{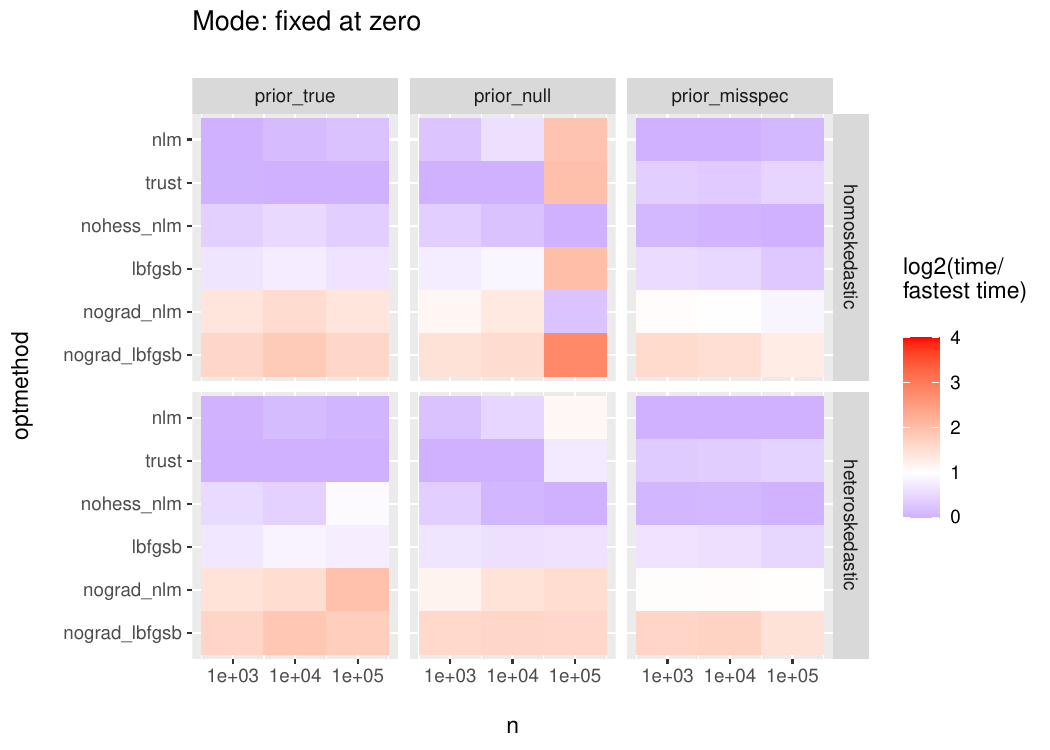}
    \includegraphics[width=.9\textwidth]{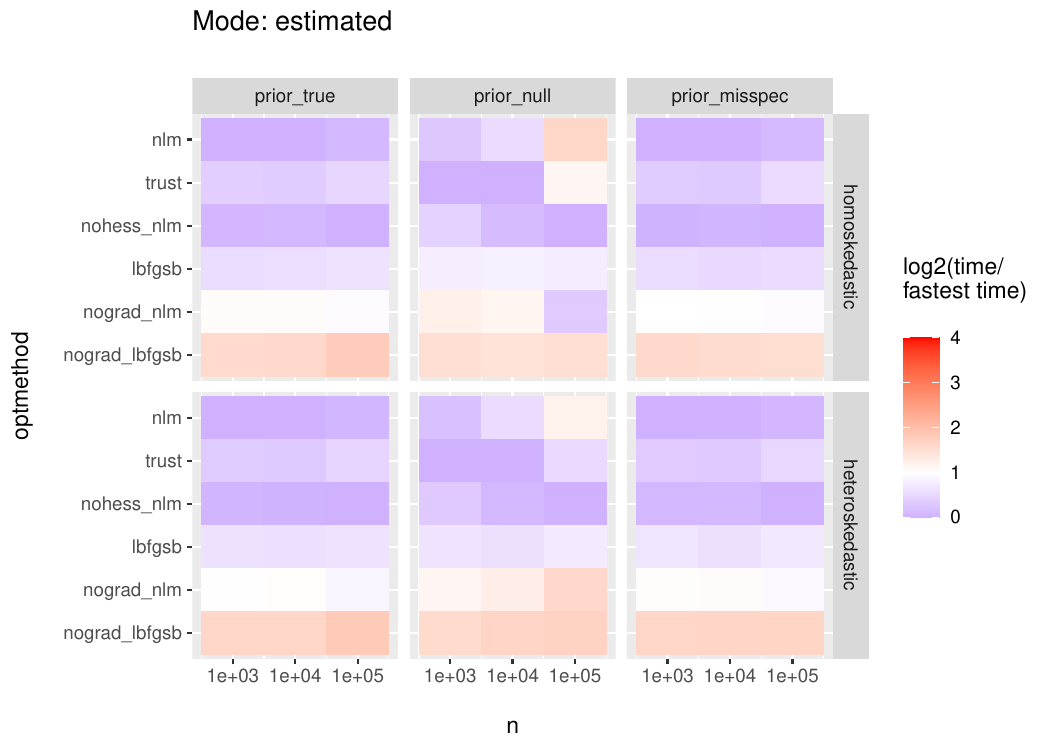}
    \caption{Timing comparisons for \code{ebnm\_point\_laplace()}. See Figure~\ref{fig:ebnm_pn} caption and text for details.}
    \label{fig:ebnm_pl}
\end{figure}

For each scenario, we ran $10^6 / n$ simulations (so, depending on $n$, 1000, 100, or 10 simulations) and compared runtimes using package \pkg{microbenchmark} \citep{microbenchmark}. All experiments were performed on a 2021 MacBook Pro with an Apple M1 Max processor and 64 GB of unified memory. Results are displayed in Figures~\ref{fig:ebnm_pn} and \ref{fig:ebnm_pl}.

In general, the methods that supplied the gradient function outperformed the methods that required all derivatives to be estimated numerically. 
Timing was similar among the four methods that do supply the gradient. However, method \code{"lbfgsb"} failed to converge in several of the \code{prior\_null} simulations (in some scenarios, up to 7\% of simulations resulted in an error). Further, the methods that supply Hessians (\code{"nlm"} and \code{"trust"}) occasionally struggled when the data-generating prior was the null distribution $\delta_0$. 
We thus recommend the default setting \code{optmethod = "nohess\_nlm"}.

\subsection{Comparisons with existing packages}
\label{sec:compare_with_existing_packages}

Next we compare the performance of \pkg{ebnm} against three packages with directly comparable functions: function \code{ebnm\_point\_laplace()} is closely related to function \code{ebayesthresh()} in the \pkg{EbayesThresh} package \citep{EbayesThresh}; function \code{ebnm\_normal\_scale\_mixture()} is modelled on function \code{ash()} in the \pkg{ashr} package  (with option \code{\nohyphens{mixcompdist} = "normal"}; \cite{ashr}) but is implemented in a much simpler manner; and function \code{ebnm\_npmle()} performs a similar task to function \code{GLmix()} in the \pkg{REBayes} package \citep{KoenkerGu}.

\begin{figure}
    \centering
    \includegraphics[width=.8\textwidth]{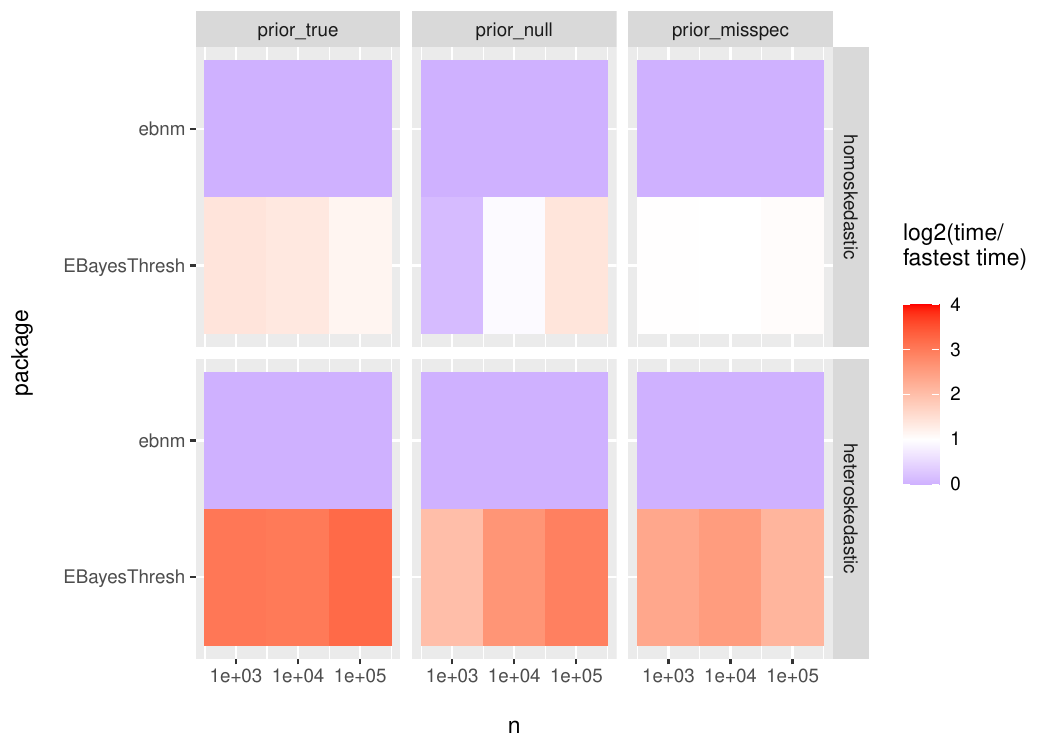}
    \caption{Timing comparisons: \code{ebnm\_point\_laplace} vs. \code{ebayesthresh}. See Figure~\ref{fig:ebnm_pn} caption and text for details.}
    \label{fig:ebayesthresh}
\end{figure}

\begin{figure}
    \centering
    \includegraphics[width=.8\textwidth]{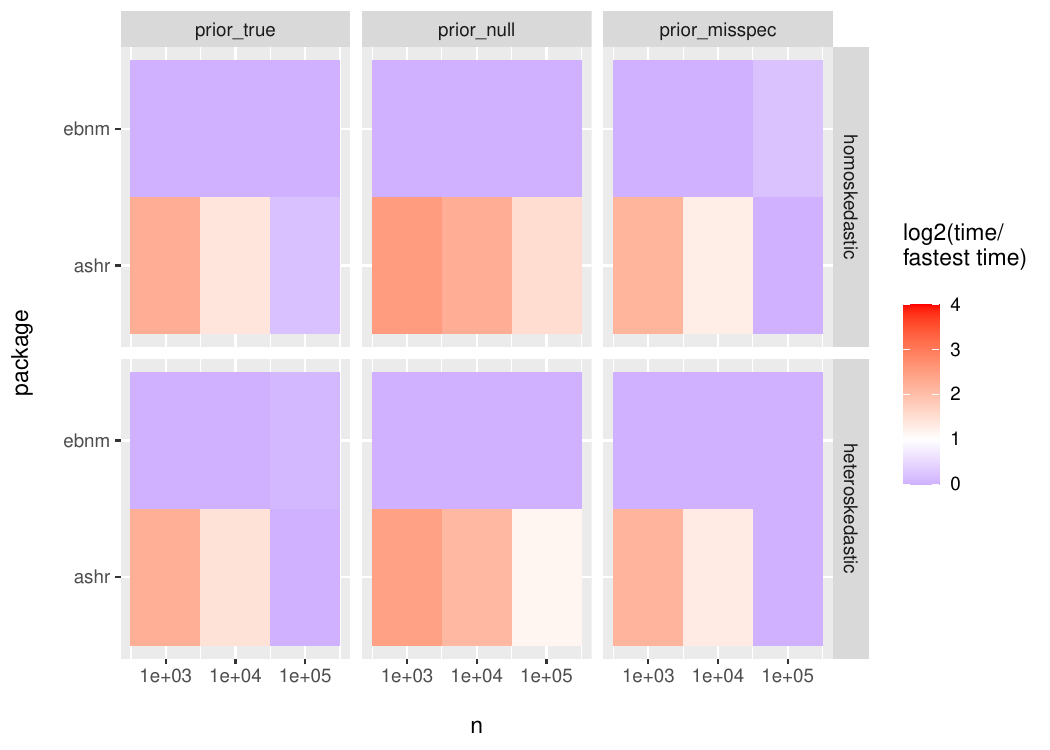}
    \caption{Timing comparisons: \code{ebnm\_normal\_scale\_mixture} vs. \code{ash}. See Figure~\ref{fig:ebnm_pn} caption and text for details.}
    \label{fig:ashr}
\end{figure}

\begin{figure}
    \centering
    \includegraphics[width=.9\textwidth]{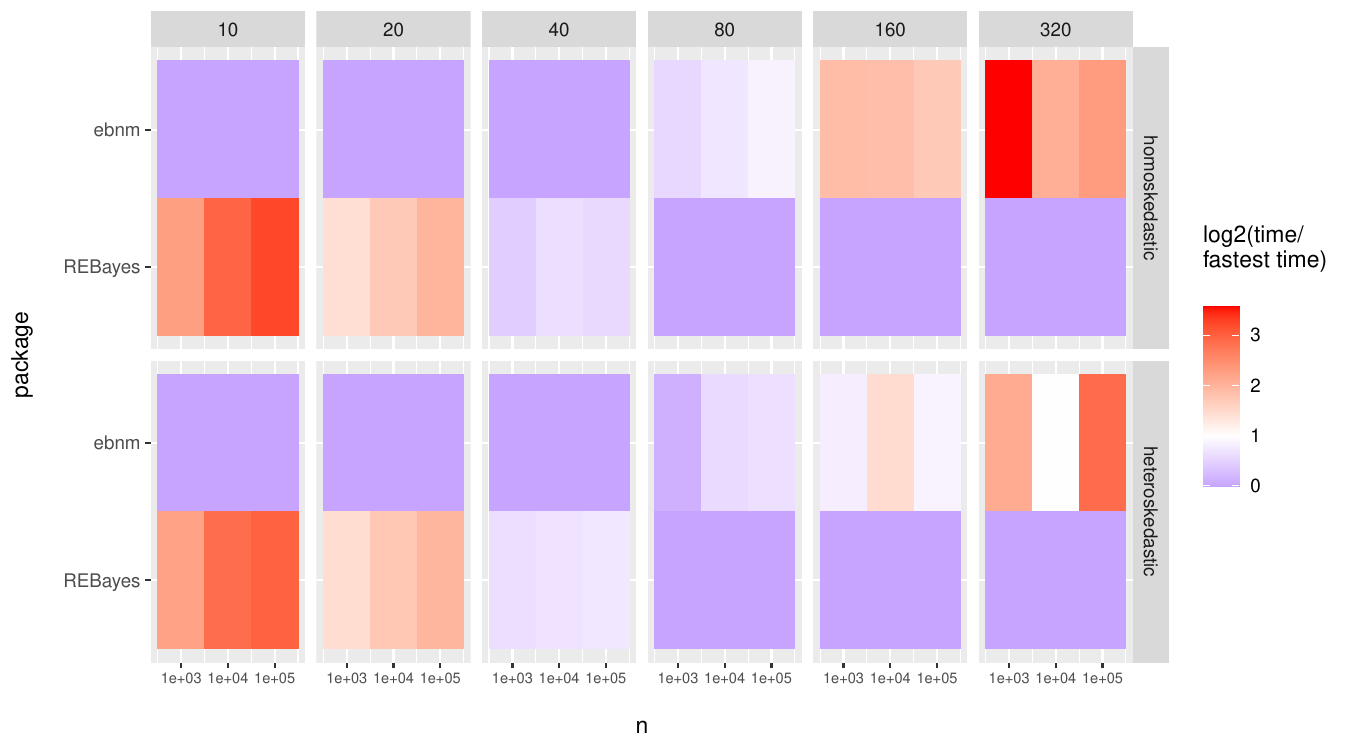}
    \caption{Timing comparisons: \code{ebnm\_npmle} vs. \pkg{REBayes}. Column headers indicate the number of grid points (mixture components) that were used to estimate the NPMLE. All simulations were from a point-Laplace distribution; the noise models (homoskedastic and heteroskedastic) are as above (see Figure~\ref{fig:ebnm_pn} caption for details). The sample sizes used were $n = 10^3$, $n = 10^4$, and $n = 10^5$.}
    \label{fig:rebayes}
\end{figure}

We ran tests for the same scenarios as Section~\ref{sec:optmethod}, with the difference that the mode is always fixed at zero (mode estimation is not possible with \pkg{EbayesThresh}). Further, since it is not possible to ``misspecify'' the prior for the family of all distributions $\mathcal{G}_{\text{npmle}}$, we only considered a single data-generating distribution (the point-Laplace), but we varied the number of grid points (mixture components) from 10 to 300. The number of simulations, \code{microbenchmark} settings, and hardware were as described in Section~\ref{sec:optmethod}. We set parameters to make outputs as similar as possible. For \pkg{EbayesThresh}, we set \code{threshrule = "mean"} and \code{universalthresh = FALSE}; for \code{ash()}, we set \code{prior = "uniform"}. Results are given in Figures~\ref{fig:ebayesthresh}--\ref{fig:rebayes}.

In some scenarios, \pkg{EbayesThresh} was nearly as fast as \code{ebnm\_point\_laplace()}, but in others it was outperformed by a full order of magnitude. Further, \pkg{ebnm} regularly found significantly better solutions than \pkg{EbayesThresh} (in terms of the final objective attained) except when the data-generating prior was the null distribution, in which cases the packages found solutions of similar quality. 

When the number of observations was small, \pkg{ebnm} was faster than \pkg{ashr} by a factor of around 2 to 4, but \code{ash} performed comparably to \code{ebnm\_normal\_scale\_mixture()} for large problems. 

Results in the comparison between \code{ebnm\_npmle()} and \pkg{REBayes} were mixed. \pkg{ebnm} was regularly faster when the number of mixture components was small (fewer than 80), while \pkg{REBayes} was consistently faster when a dense grid was used (80 or more  components). 
According to the theory developed in \cite{WillwerscheidDiss}, 80 components should be ``good enough'' for homoskedastic observations when
\begin{equation}
n^{1/4} \left( \frac{\text{range}(x)}{s} \right) \le 80 \sqrt{8}.
\end{equation}
For example, if the number of observations $n = 10^4$, then 80 components should suffice as long as the studentized range
\begin{equation}
\frac{\max(x) - \min(x)}{s} \le 8 \sqrt{8} \approx 22.6.
\end{equation}

\section{Background on ``weighted on-base averages''}
\label{sec:woba}

A longstanding tradition in empirical Bayes research is to include an
analysis of batting averages using data from Major League Baseball
(see, for example, \citealt{BrownBaseball, JiangZhangBaseball,
  GuKoenkerBaseball}). Until recently, batting averages were the most
important measurement of a hitter's performance, with the prestigious
yearly ``batting title'' going to the hitter with the highest
average. However, with the rise of baseball analytics, metrics that
better correlate to teams' overall run production have become
increasingly preferred.  One such metric is wOBA (``weighted on-base
average''), which is both an excellent measure of a hitter's offensive
production and, unlike competing metrics such as MLB's xwOBA
\citep{xwoba} or Baseball Prospectus's DRC+ \citep{drcplus}, can be
calculated using publicly available data and methods.

Initially proposed by \cite{Tango}, wOBA assigns values 
(``weights'') to hitting outcomes according to how much the outcome
contributes on average to run production. For example, while
batting average treats singles identically to home runs, wOBA
gives a hitter more than twice as much credit for a home
run.\footnote{Weights are updated from year to year, but wOBA weights
  for singles have remained near 0.9 for the last several decades,
  while weights for home runs have hovered around 2.0
  \citep{fgguts}.}

Given a vector of wOBA weights $\mathbf{w}$, hitter $i$'s wOBA is the
weighted average
\begin{equation}
x_i \colonequals \mathbf{w}^\top \mathbf{z}^{(i)} / n_i,
\end{equation}
where $\mathbf{z}^{(i)} = (z_1^{(i)}, \ldots, z_7^{(i)})$ tallies
  outcomes (singles, doubles, triples, home runs, walks,
  hit-by-pitches and outs) over the hitter's $n_i$ plate appearances
  (PAs). Modeling hitting outcomes as i.i.d.
\begin{equation} 
\mathbf{z}^{(i)} \sim \text{Multinomial}(n_i, {\bm\pi}^{(i)}),
\label{eqn:woba_multinom}
\end{equation}
where ${\bm\pi}^{(i)} = (\pi_1, \ldots, \pi_7^{(i)})$ is the vector of
``true'' outcome probabilities for hitter $i$, we can regard $x_i$ as
a point estimate for the hitter's ``true wOBA skill'',
\begin{equation} 
\theta_i \colonequals \mathbf{w}^\top {\bm\pi}^{(i)}.
\end{equation}
Standard errors for the $x_i$'s can be estimated as
\begin{equation}
s_i^2 = \mathbf{w}^\top \hat{\bm\Sigma}^{(i)} \mathbf{w}/n_i,
\end{equation}
where $\hat{\bm\Sigma}^{(i)}$ is the estimate of the covariance matrix
for the multinomial model \eqref{eqn:woba_multinom} obtained by
setting ${\bm\pi} = \hat{\bm\pi}$,\footnote{To deal with small
sample sizes, we conservatively lower bound each standard error by
the standard error that would be obtained by plugging in
league-average event probabilities $\hat{\bm\pi}_{\mathrm{lg}} =
\sum_{i=1}^N \mathbf{z}^{(i)}/ \sum_{i=1}^N n_i$, where $N$ is the
number of hitters in the data set.} where
\begin{equation} 
\hat{\bm\pi}^{(i)} = \mathbf{z}^{(i)}/n_i.
\end{equation}

The relative complexity of wOBA makes it well suited for analysis via
\pkg{ebnm}. With batting average, a common approach is to obtain
empirical Bayes estimates using a beta-binomial model (see, for
example, \citealt{robinson}). With wOBA, one can estimate hitting
outcome probabilities by way of a Dirichlet-multinomial model;
alternatively, one can approximate the likelihood as normal and fit an
EBNM model directly to the observed wOBAs. We take the latter
approach.

\end{document}